\documentclass{article}
\usepackage{flafter}                    
\usepackage{amsmath}                    
\usepackage{amssymb}
\usepackage{mathrsfs}
\usepackage{eufrak}
\usepackage{eucal}
\usepackage[dvips]{graphicx}            
\newcommand{\hil}{\mathcal H}
\newcommand{\la}{\langle}
\newcommand{\ra}{\rangle}
\newcommand{\f}{\varphi}
\newcommand{\F}{\Phi}

\newcommand{\e}{\varepsilon}
\newcommand{\ka}{\varkappa}
\newcommand{\om}{\omega}
\newcommand{\Om}{\Omega}

\newcommand{\nR}{\mathbb R}
\newcommand{\nC}{\mathbb C}
\newcommand{\nN}{\mathbb N}

\newcommand{\nT}{\mathbb T}
\newcommand{\nZ}{\mathbb Z}
\newcommand{\sis}[3]{\la#1|#2\ra_{#3}}

\newcommand{\suor}[1]{\int_{#1}^{\oplus}}

\newcommand{\vek}[1]{\mathbf{#1}}
\newcommand{\veg}[1]{\boldsymbol{#1}}

\newcommand{\mr}[1]{\mathrm{#1}}
\newcommand{\mc}[1]{\mathcal{#1}}
\newcommand{\mf}[1]{\mathfrak{#1}}
\newtheorem{prop}{Proposition}[section]

\newtheorem{theor}{Theorem}[section]

\newtheorem{rem}{Remark}[section]
\title{Extreme Covariant Quantum Observables in the Case of an Abelian Symmetry Group and a Transitive Value Space}
\author{Erkka Theodor Haapasalo\footnote{ethaap@utu.fi} \\ Juha-Pekka Pellonp\"a\"a\footnote{juhpello@utu.fi}}

\begin{document}
\maketitle
\begin{center}
Turku Centre for Quantum Physics, Department of Physics and Astronomy, University of Turku, FI-20014 Turku, Finland
\end{center}
\section*{Abstract}

We represent quantum observables as POVMs (normalized positive operator valued measures) and consider convex sets of observables which are covariant with respect to a unitary representation of a locally compact Abelian symmetry group $G$. The value space of such observables is a transitive $G$-space. We characterize the extreme points of covariant observables and also determine the covariant extreme points of the larger set of all quantum observables. The results are applied to position, position difference and time observables.\\[5pt]
\emph{Keywords:} positive operator valued measure, POVM, convex set, extreme point, extreme observable, covariant observable, Abelian group, position observable, time observable\\[5pt]
\emph{PACS numbers:} 03.65.Ta, 03.67.-a

\section*{Introduction}

In the quantum mechanical description, observables are represented as normalized positive operator valued measures (POVMs). Observables with the same value space form a convex set. The extreme points of this set (or any of its relevant convex subset) represent observables whose measurements involve no arbitrariness caused by mixing of different measurement schemes. In this paper we concentrate on the problem of determining the extreme observables covariant with respect to a locally compact Abelian symmetry group $G$. The value space $\Om$ of such observables is a transitive $G$-space so that the characterization of extreme covariant observables is a generalization of the characterization of Holevo {\it et al} \cite{ho3} where $\Om=G$ is assumed. Since many important observables are described by covariant POVMs, characterizing extreme covariant observables is of great physical interest.

Several different extremality conditions for quantum observables are already given in literature: In \cite{pel} a complete characterization of extreme quantum observables is obtained; see also \cite{sys}. The problem of determining {\it covariant} extreme points has been studied, in the case of a compact symmetry group and an arbitrary transitive value space in \cite{car} and in the case of a general unimodular separable type I locally compact symmetry group which is also the value space in \cite{ho3}. Similar characterizations for covariant extreme POVMs in the finite-dimensional case are given in \cite{dar1}.

The structure of observables that are covariant with respect to a locally compact second countable (Hausdorff) Abelian group $G$ and whose value space is an arbitrary transitive space of $G$ has been fully determined by Cassinelli {\it et al} \cite{cas}; this paper builds upon their result. In the next section we will give a mathematical description of (covariant) quantum observables and, in Section \ref{cdt}, we restate the result of \cite{cas} on covariant observables and characterize the extreme points in the set of covariant observables. We also address the problem of defining the covariant POVMs which are extreme points in the set of all quantum observables; a characterization of such observables is given in Section \ref{cdt}. In the following sections we study some physical examples -- time, position and position difference observables -- where we apply the results obtained in the preceding sections. For instance, we show that the canonical covariant time observable is extreme in the set of all observables.
All proofs of the theoretical part are given in Appendix A.

\section{Preliminaries}\label{prel}

Let $\nN:=\{0,1,\ldots\}$. Suppose that $(\Om,\,\mc A)$ is a measurable space (where $\mc A$ is a $\sigma$-algebra consisting of subsets of a set $\Om$) and that $\hil$ is a (complex) Hilbert space, and $\mc L(\hil)$ the algebra of bounded operators on $\hil$. Let $M:\mc A\to\mc L(\hil)$ be an {\it operator (valued) measure,} i.e.\ weakly $\sigma$-additive mapping: for any disjoint sequence $(B_k)_{k\in\nN}\subset\mc A$ the condition
$
\sis{\f}{M(\cup_{k\in\nN}B_k)\psi}{}=\sum_{k\in\nN}\sis{\f}{M(B_k)\psi}{}
$
holds for all $\f,\,\psi\in\hil$. We call an operator measure $M$ \emph{positive} if for all $B\in\mc A$,  $M(B)\geq 0$ i.e.\ $\sis{\f}{M(B)\f}{}\geq0$ for all $\f\in\hil$. An operator measure $M$ is \emph{projection valued} if its range consists of projections: $M(B)^2=M(B)^*=M(B)$ for all $B\in\mc A$. An operator measure is \emph{normalized} if $M(\Om)=I=I_\hil$ (the identity operator of $\hil$). Normalized positive operator valued measures (POVMs) are identified with {\it observables.} Normalized projection valued measures (PVMs) are often called \emph{sharp observables}. The set of POVMs $M:\mc A\to\mc L(\hil)$ is denoted by $\mc O(\mc A,\,\hil)$ and the corresponding set of PVMs is denoted by $\Sigma(\mc A,\,\hil)$.

The set $\mc O(\mc A,\,\hil)$ is convex: We may take any observables $M_1,\,M_2\in\mc O(\mc A,\,\hil)$ and any $0\leq t\leq1$ and define the combined observable $tM_1+(1-t)M_2$ by
\begin{equation}\label{conv}
\big(tM_1+(1-t)M_2\big)(B)=tM_1(B)+(1-t)M_2(B),\qquad B\in\mc A.
\end{equation}
A convex combination of the form \eqref{conv} can be seen as a mixing or randomization of measuring procedures of the observables $M_1$ and $M_2$. We may then ask what are the extreme elements of the set $\mc O(\mc A,\,\hil)$. As usual, an observable $M\in\mc O(\mc A,\,\hil)$ is \emph{extreme}, $M\in\mr{Ext}\big(\mc O(\mc A,\,\hil)\big)$, when from the condition $M=tM_1+(1-t)M_2$ for any $M_1,\,M_2\in\mc O(\mc A,\,\hil)$ and $0<t<1$ it follows that $M=M_1=M_2$. We also study convex subsets $\mf M\subset\mc O(\mc A,\,\hil)$ and their extreme points. Similarly, an extreme observable $M\in\mr{Ext}(\mf M)$ cannot be obtained as a (nontrivial) combination like \eqref{conv} with $M_1,\,M_2\in\mf M$; this means that the measurement of $M$ involves no redundancy caused by combining different measurements of observables from the class $\mf M$. It is a direct consequence of \cite[Lemma 2.3]{dav}, that spectral measures are extreme points of $\mc O(\mc A,\,\hil)$ and thus $\Sigma(\mc A,\,\hil)\cap\mf M\subset\mr{Ext}(\mf M)$ when $\mf M\subset\mc O(\mc A,\,\hil)$ is a convex set of observables. Note that an extreme point of a convex set $\mf M\subset\mc O(\mc A,\,\hil)$ is not necessarily projection valued.

If $\mf M\cap\Sigma(\mc A,\,\hil)=\emptyset$, a question arises whether the convex set $\mf M\subset\mc O(\mc A,\,\hil)$ contains any extreme points. Let us assume that $\Om$ is a locally compact Hausdorff space and $\hil$ is a separable Hilbert space. Following \cite[Section IV.A]{car}  one can define  a locally convex topology of the space of operator measures on $\mc B(\Om)$ (here $\mc B(\Om)$ is the $\sigma$-algebra of Borel sets of $\Om$) acting on $\hil$ and show that the set of observables $\mc O(\Om,\,\hil):=\mc O\big(\mc B(\Om),\,\hil\big)$ is a compact subset of this space. The Krein-Milman theorem states that the set of extreme points of any closed and convex subset $\mf M$ of observables is non-empty and that $\mf M$ coincides with the closure of the convex hull of Ext$(\mf M)$.

In the sequel we concentrate on a special class of observables; covariant observables. Let us assume that $G$ is a topological group 
with a strongly continuous unitary representation $U$ acting on a separable Hilbert space $\hil$. Let $\Om$ be a topological space with a transitive continuous $G$-action $G\times\Om\ni(g,\om)\mapsto g\cdot\om\in\Om$. Suppose that we have an observable $M\in\mc O(\Om,\,\hil)$ that obeys the {\it covariance condition}
\begin{equation}\label{covar}
U(g)M(B)U(g)^*=M(g\cdot B)
\end{equation}
for all $g\in G$ and $B\in\mc B(\Om)$. We call these observables $(\Omega,U)$-\emph{covariant} or briefly \emph{covariant} if there is no possibility of confusion.
We denote the set of $(\Omega,U)$-covariant observables of $\mc O(\Om,\,\hil)$ by $\mc O_U(\Om,\,\hil)$ and call the set $\mc O_U(\Om,\,\hil)$ as a \emph{covariance structure} determined by the representation $U$ and the $G$-space $\Om$. Note that covariance structures $\mc O_U(\Om,\,\hil)$ are convex subsets of $\mc O(\Om,\,\hil)$. It can also be easily shown that covariance structures are closed in the sense of \cite{car} so that one can apply Krein-Milman theorem as above.

In this paper, we are especially interested in the covariance structures $\mc O_U(\Om,\,\hil)$ where $U$ is a strongly continuous representation of a locally compact second countable Abelian group $G$ which is Hausdorff. In this case the group $G$ is $\sigma$-compact and any transitive $G$-space $\Om$ is homeomorphic to a (left) coset group $G/H$ for some closed subgroup $H\leq G$ \cite{fol}. Hence, let $\Om=G/H$ with $H\leq G$ closed, and let $[g]=gH$ denote the coset of $\Om$ determined by $g\in G$. Suppose that $M\in\mc O(\Om,\,\hil)$ is a covariant POVM. Cattaneo \cite{cat} has shown that there is a minimal covariant Naimark dilation of $M$, i.e.\ there is a Hilbert space $\mc K$, a representation $\tilde U_0$ of $G$ in $\mc K$, an isometry $V_0:\hil\to\mc K$ and a PVM $P\in\mc O_{\tilde U_0}(\Om,\,\mc K)$ such that vectors $P(B)V_0\f$, $B\in\mc B(\Om)$, $\f\in\hil$, span a dense subspace of $\mc K$ and
\begin{equation}
\begin{array}{lcll}
V_0M(B)&=&P(B)V_0,&B\in\mc B(\Om),\nonumber\\
V_0U(g)&=&\tilde U_0(g)V_0,&g\in G.\nonumber
\end{array}
\end{equation}

Mackey's imprimitivity theorem \cite[Theorem 6.31]{fol}, \cite{ma1} states that there is a representation $\pi$ of the subgroup $H$ in a Hilbert space $\hil_\pi$, a $G$-invariant (Haar) measure $\mu:\mc B(\Om)\to[0,\infty]$ and a Hilbert space $\mf H$ of functions $f:G\to\hil_\pi$ such that
\begin{itemize}
\item the function $G\ni g\mapsto\sis{\f}{f(g)}{}$ is Borel for all $\f\in\hil_\pi$,
\item $f(gh)=\pi(h)f(g)$ for all $g\in G$ and $h\in H$ and
\item $\int_\Om\|f(g)\|^2\,d\mu([g])<\infty$.
\end{itemize}
The inner product of the space $\mf H$ is given by
$$
\sis{f_1}{f_2}{}=\int_\Om\sis{f_1(g)}{f_2(g)}{}\,d\mu([g]),\qquad f_1,\,f_2\in\mf H.
$$
Let us define a PVM $\tilde P\in\Sigma(\Om,\,\mf H):=\Sigma\big(\mc B(\Om),\,\mf H\big)$ and a strongly continuous representation $\tilde U$ of $G$ in $\mf H$ such that
\begin{eqnarray}
\big(\tilde P(B)f\big)(g)&=&\chi_B([g])f(g),\\
\big(\tilde U(g')f\big)(g)&=&f(g'g)
\end{eqnarray}
for all $B\in\mc B(\Om)$, $f\in\mf H$ and $g,\,g'\in G$. The representation $\tilde U$ is the \emph{representation induced from $\pi$} and it is often denoted by $\mr{ind}_H^G(\pi)$. The function $\chi_B:\Om\to\{0,\,1\}$ is the characteristic function or indicator of $B$. In addition, there is a unitary mapping $V_1:\mc K\to\mf H$ intertwining $P$ with $\tilde P$ and respectively $\tilde U_0$ with $\tilde U$, and hence the isometry $V:=V_1V_0$ intertwines $M$ with $\tilde P$ and $U$ with $\tilde U$, i.e.
\begin{equation}
\begin{array}{lcll}
VM(B)&=&\tilde P(B)V,&B\in\mc B(\Om),\\
VU(g)&=&\tilde U(g)V,&g\in G.
\end{array}
\end{equation}
The triple $(\tilde P,\,\tilde U,\,\Om)$ which is unique up to unitary equivalence is called the \emph{canonical system of imprimitivity} corresponding the triple $(M,\,U,\,\Om)$. The results of \cite{cat} are also valid when $G$ is not Abelian. Using this result Cassinelli {\it et al} have characterized the covariance structure $\mc O_U(\Om,\,\hil)$ in case of an Abelian symmetry group \cite{cas}. We summarize these results in the following section. For basic results in harmonic analysis, we refer to \cite{fol} and \cite{dix}.

\section{Extremality Conditions of Covariant Observables}\label{cdt}

For any topological space $X$, we let $C_c(X)$ denote the space of continuous compactly supported functions $f:\,X\to\mathbb C$. 
For any measure $\lambda:\,\mathcal B(X)\to[0,\infty]$ and any Hilbert space $\mathcal M$ we let $L^2(X,\lambda;\mathcal M)$ denote the Hilbert space of (equivalence classes of) square integrable functions $X\to\mathcal M$ and we let $L^2(X,\lambda):=L^2(X,\lambda;\mathbb C)$. Moreover, we denote the Banach space of  $\lambda$-integrable functions $f:X\to\nC$ by $L^1(X,\,\lambda)$ and $(L^1\cap L^2)(X,\lambda):=L^1(X,\lambda)\cap L^2(X,\lambda)$. If $X\subset\nR^n$ and $\lambda$ is the (restricted and possibly scaled) Lebesgue measure, we drop $\lambda$ out from the above notations.

We also need the concept of a direct integral of Hilbert spaces. Suppose that $X$ is a nonempty set with a $\sigma$-algebra $\mc A$ of its subsets. Suppose that $\mu:\mc A\to[0,\infty]$ is a measure and $(\hil_x,\,\sis{\cdot}{\cdot}{x})$ is a separable Hilbert space for all $x\in X$. Suppose that there are vector fields $X\ni x\mapsto e_k(x)\in\hil_x$, $k\in\nN$, such that $e_k(x)$, $k\in\nN$, generate the Hilbert space $\hil_x$. We may even assume that the vectors $e_k(x)$, $k<\dim{(\hil_x)}+1$, form an orthonormal basis of $\hil_x$ for every $x\in X$. We call a section $\f\in\prod_{x\in X}\hil_x$ measurable, if the functions $X\ni x\mapsto\sis{e_k(x)}{\f(x)}{x}\in\nC$ are $\mc A$-measurable. Let the direct integral $\suor{X}\hil_x\,d\mu(x)$ be the Hilbert space of $\mu$-square-integrable sections equipped with the inner product
$$
(\f,\psi)\mapsto\sis{\f}{\psi}{}=\int_X\sis{\f(x)}{\psi(x)}{x}\,d\mu(x).
$$
Clearly $L^2$-spaces are of the direct integral form. We call an operator $A$ of $\suor{X}\hil_x\,d\mu(x)$ {decomposable with components} $A(x)$ on $\hil_x$, $x\in X$, if $(A\f)(x)=A(x)\f(x)$ for all $\f\in\suor{X}\hil_x\,d\mu(x)$ and $x\in X$. The decomposable operator is denoted by
$$
A=\suor{X}A(x)\,d\mu(x)
$$
and it is bounded iff $\|A\|:=\mr{ess}\,\sup_{x\in X}\|A(x)\|<\infty$. If the measurable space $(X,\,\mc A)$ is discrete, the direct integral $\suor{X}\hil_x\,d\mu(x)$ reduces to a direct sum.

{\it For the rest of this paper we assume that}
\begin{itemize}
\item $G$ is a locally compact second countable Abelian group which is Hausdorff,
\item $U$ is a strongly continuous unitary representation of $G$ in a separable (complex) Hilbert space $\hil$,
\item $\Om=G/H$ where $H\leq G$ is a closed subgroup,
\item $\mu$ is an (essentially unique) $G$-invariant Borel measure on $\Om$,
\item $\hat G$ is the character group, the representation dual, of $G$,
\item $H^\perp$ is the annihilator of $H$, i.e.\ the subgroup of those $\eta\in\hat G$ such that $\la h,\eta\ra=1$ for all $h\in H$.
\end{itemize}
Here we have denoted $\la g,\gamma\ra:=\gamma(g)\in\nT$ for all $g\in G$ and $\gamma\in\hat G$.

Since $G$ is of type I, there is a Borel measure $\nu_U$ on character group $\hat G$ which is finite on compact sets and a measurable field of Hilbert spaces $\hil_\gamma$, $\gamma\in\hat G$, such that the Hilbert space $\hil$ can be given as a direct integral
\begin{equation}\label{suora}
\hil=\suor{\hat G}\hil_\gamma\,d\nu_U(\gamma)
\end{equation}
where the representation $U$ operates diagonally, i.e.\ $\big(U(g)\f\big)(\gamma)=\la g,\gamma\ra\f(\gamma)$ for all $g\in G$, $\f\in\hil$ and $\gamma\in\hat G$.

If there are observables covariant with respect to the representation $U$, i.e.\ $\mc O_U(\Om,\,\hil)\neq\emptyset$, we may dilate the generalized system of imprimitivity $(M,\,U,\,\Om)$ into a canonical system of imprimitivity induced from a strongly continuous unitary representation $\pi$ of the subgroup $H$ operating in a Hilbert space $\hil_\pi$ \cite{cat}. Thus there is a Borel measure $\nu$ on $\hat H\simeq\hat G/H^\perp$ which is finite on compact sets that gives the direct integral decomposition
\begin{equation}\label{suora2}
\hil_\pi=\suor{\hat G/H^\perp}\hil_{[\gamma]}\,d\nu([\gamma])
\end{equation}
where the representation $\pi$ operates diagonally, i.e.\ $\big(\pi(h)\f\big)([\gamma])=\la h,\gamma\ra\f([\gamma])$ for all $h\in H$, $\f\in\hil_\pi$ and $\gamma\in\hat G$. 
Here $[\gamma]=\gamma+H^\perp$ is the coset in $\hat G/H^\perp$ determined by $\gamma\in\hat G$. 

Pick a Haar measure $d\eta$ for $H^\perp$ such that the Fourier-Plancherel transformation $\mc F:L^2(\Om,\mu)\to L^2(H^\perp,d\eta)$ defined by
$$
(\mc Ff)(\eta)=\int_\Om\la g,\eta\ra f([g])\,d\mu([g]),\qquad f\in(L^1\cap L^2)(\Om,\mu),\quad\eta\in H^\perp,
$$
is unitary. 
Define a continuous positive functional $L:C_c(\hat G)\to\nC$,
\begin{equation}\label{lfnc}
L(f)=\int_{\hat G/H^\perp}\int_{H^\perp}f(\gamma+\eta)\,d\eta\,d\nu([\gamma])=\int_{\hat G}f(\gamma)\,d\tilde\nu(\gamma)
\end{equation}
for all $f\in C_c(\hat G)$ where the measure $\tilde\nu:\mc B(\hat G)\to[0,\infty]$ given by the Riesz representation theorem is finite on compact sets. The measure $\tilde\nu$ can be viewed as a `lift' of $\nu$ with respect to the `fibration' $\hat G\to\hat G/H^\perp,\,\gamma\mapsto[\gamma]$.

The following result has been obtained in \cite{cas}.
\begin{prop}\label{prop1}
Let $\nu_U$ be the measure giving the direct integral decomposition \eqref{suora} in which $U$ acts diagonally. The covariance structure $\mc O_U(\Om,\,\hil)$ is non-empty if and only if there is a strongly continuous unitary representation $\pi$ of the subgroup $H$ in $\hil_\pi$ such that there is an isometry $W_0:\hil\to\mf H$ from the space of the representation $U$ to the space $\mf H$ of the induced representation $\tilde U:=\mr{ind}_H^G(\pi)$ intertwining $U$ and $\tilde U$, i.e.\ $W_0U(g)=\tilde U(g)W_0$ for all $g\in G$. Moreover, in this case, the measure $\nu_U$ is absolutely continuous with respect to the lifted measure $\tilde\nu$ of \eqref{lfnc} derived from the measure $\nu$ of \eqref{suora2}.
\end{prop}
Without restricting generality (see Remark \ref{tih} in Appendix \ref{aa}) {\it we simply assume that} 
$$
\boxed{\nu_U=\tilde\nu.}
$$

Fix an infinite-dimensional Hilbert space $\mc M$. Suppose that $\f\in\hil=\suor{\hat G}\hil_\gamma\,d\tilde\nu(\gamma)$ and $[\gamma]\in\hat G/H^\perp$. Let us define
$$
\|\f_{[\gamma]}\|_1:=\int_{H^\perp}\|\f(\gamma+\eta)\|\,d\eta
$$
whenever the integral exists. One can easily check that the set
\begin{equation}
\mc D:=\left\{\f\in\hil\,\Bigg|\,\int_{\hat G/H^\perp}\|\f_{[\gamma]}\|_1^2\,d\nu([\gamma])<\infty\right\}
\end{equation}
is a linear subspace of $\hil$.

Suppose that $W:\hil\to L^2(\hat G,\,\tilde\nu;\,\mc M)$ is a decomposable isometry, i.e $(W\f)(\gamma)=W(\gamma)\f(\gamma)$ for all $\f\in\hil$ and $\gamma\in\hat G$ where $W(\gamma):\hil_\gamma\to\mc M$ is an isometry. Define an operator $\mf W:\mc D\to L^2(\hat G/H^\perp,\nu;\mc M)$ such that
\begin{equation}\label{mfw}
(\mf W\f)([\gamma])=\int_{H^\perp} W(\gamma+\eta)\f(\gamma+\eta)\,d\eta\in\mc M
\end{equation}
for all $\f\in\mc D$ and $[\gamma]\in\hat G/H^\perp$. Let $\f\in\mc D$. The operator $\mf W$ is well defined and
$$
\|\mf W\f\|^2\leq\int_{\hat G/H^\perp}\|\f_{[\gamma]}\|_1^2\,d\nu([\gamma])<\infty
$$
for all $\mc D$.
We may reformulate the main result of \cite{cas} in the following form. The original formulation of \cite{cas} (Theorem \ref{castheor}) and the proofs of the theorems below are in Appendix \ref{aa}. Results concerning covariant PVMs are essentially from \cite{hoa}.

\begin{theor}\label{theor1}
For any $M\in\mc O_U(\Om,\,\hil)$ there is a decomposable isometry $W:\hil\to L^2(\hat G,\,\tilde\nu;\,\mc M)$ with isometric components $W(\gamma):\hil_\gamma\to\mc M$, $\gamma\in\hat G$, such that
\begin{equation}\label{kolm}
\sis{\f}{M(B)\psi}{}=\int_B\sis{\mf WU(g)^*\f}{\mf WU(g)^*\psi}{}\,d\mu([g])
\end{equation}
for all $\f,\,\psi\in\mc D$  and $B\in\mathcal B(\Omega)$. The operator $\mf W:\mc D\to L^2(\hat G/H^\perp,\,\nu;\,\mc M)$ is obtained from $W$ as above. On the other hand, given a decomposable isometry $W:\hil\to L^2(\hat G,\,\tilde\nu;\,\mc M)$, Equation \eqref{kolm} defines an observable $M\in\mc O_U(\Om,\,\hil)$. The intersection $\mc O_U(\Om,\,\hil)\cap\Sigma(\Om,\,\hil)$ is non-empty if and only if the function $\hat G\ni\gamma\mapsto n(\gamma):=\dim{(\hil_\gamma)}$ is essentially constant on (almost) every coset $[\gamma]\in\hat G/H^\perp$. Furthermore, an observable $M\in\mc O_U(\Om,\,\hil)$ is a PVM if and only if the mapping $W(\gamma_2)^*W(\gamma_1):\hil_{\gamma_1}\to\hil_{\gamma_2}$ is unitary for a.a.\ $[\gamma]\in\hat G/H^\perp$ and a.a.\ $\gamma_1,\,\gamma_2\in[\gamma]$.
\end{theor}

Let us denote by $\vek H$ the closure of the image space $\mf W(\mc D)$ in $L^2(\hat G/H^\perp,\,\nu;\,\mc M)$. It is shown in the Appendix \ref{aa} that there is a measurable field of Hilbert spaces $\hat G/H^\perp\ni[\gamma]\mapsto\vek H_{[\gamma]}\subset\mc M$ giving the direct integral decomposition
\begin{equation}\label{suorah}
\vek H=\suor{\hat G/H^\perp}\vek H_{[\gamma]}\,d\nu([\gamma]).
\end{equation}
Note that $W(\gamma)\hil_\gamma\subset\vek H$ for a.a.\ $\gamma\in\hat G$
and we say that \eqref{kolm} is the {\it minimal Kolmogorov decomposition of $M$,} see \cite{ho3,pel}.
We are ready to characterize the extreme points of $\mc O_U(\Om,\,\hil)$.

\begin{theor}\label{theor2}
Suppose that $M\in\mc O_U(\Om,\,\hil)$ and the decomposable isometry $W$ is as in Theorem \ref{theor1}. Let us keep the notations introduced in this section. The observable $M$ is extreme in the covariance structure, $M\in\mr{Ext}(\mc O_U(\Om,\,\hil))$, if and only if there is no non-zero decomposable operator $A\in\mc L(\vek H)$,
\begin{equation}\label{suoraA}
A=\suor{\hat G/H^\perp}A_{[\gamma]}\,d\nu([\gamma]),\qquad A_{[\gamma]}\in\mc L(\vek H_{[\gamma]}),\quad[\gamma]\in\hat G/H^\perp,
\end{equation}
such that 
\begin{equation}\label{extehto}
W(\gamma')^*A_{[\gamma]}W(\gamma')=0 
\end{equation}
for a.a.\ $[\gamma]\in\hat G/H^\perp$ and a.a.\ $\gamma'\in[\gamma]$.
\end{theor}

An observable $M\in\mc O_U(\Om,\,\hil)$ that is extreme in $\mc O(\Om,\,\hil)$ is, of course, also extreme in the covariance structure $\mc O_U(\Om,\,\hil)$. Typically the set of convex extreme points of $\mc O_U(\Om,\,\hil)$ is larger than the set of extreme points of $\mc O(\Om,\,\hil)$ that are contained in $\mc O_U(\Om,\,\hil)$, i.e.\ an observable that is extreme in $\mc O_U(\Om,\,\hil)$ need not be extreme in $\mc O(\Om,\,\hil)$. Next we characterize covariant observables that are extreme also in $\mc O(\Om,\,\hil)$. Note that there are Borel-measurable sections $s:\Om\to G$ for the quotient projection $G\ni g\mapsto[g]\in\Om$ \cite[lemma 3]{bag}.

\begin{theor}\label{theor4}
An observable $M\in\mc O_U(\Om,\,\hil)$ is extreme in $\mc O(\Om,\,\hil)$ if and only if there is no non-zero decomposable operator $D\in\mc L\big(L^2(\Om,\mu;\vek H)\big)$ with components $D(\om)\in\mc L(\vek H)$ for all $\om\in\Om$ such that for some measurable section $s:\Om\to G$
\begin{eqnarray}\label{kovextehto}
0&=&\int_\Om\sis{\mf W(U\circ s)(\om)^*\f}{D(\om)\mf W(U\circ s)(\om)^*\psi}{}d\mu(\om)\\
&=&\int_\Om\int_{\hat G/H^\perp}\int_{H^\perp}\int_{H^\perp}\la\zeta-\xi,s(\om)\ra\sis{W(\gamma+\zeta)\f(\gamma+\zeta)}{D(\om)W(\gamma+\xi)\psi(\gamma+\xi)}{}\,d\xi\,d\zeta\,d\nu([\gamma])\,d\mu(\om)\nonumber
\end{eqnarray}
for all $\f,\,\psi\in\mc D$.
\end{theor}

\begin{rem}\rm 
Let $M\in\mc O_U(\Om,\,\hil)$ with the corresponding space $\vek H$.
We say that $\dim\vek H$ is the {\it rank of} $M$. If the rank of $M$ is $1$, i.e.\
$\vek H\simeq\nC$ also $\hil_\gamma\simeq\nC$ (or $\hil_\gamma=\{0\}$) for a.a.\ $\gamma\in\hat G$. Denoting $\Lambda=\{\gamma\in\hat G\,|\,\dim\hil_\gamma=1\}$ we see that $\hil\simeq L^2\big(\Lambda,\tilde\nu|_{\mc B(\Lambda)}\big)$. Furthermore, there must be a $[\gamma_0]\in\hat G/H^\perp$ such that $\nu\big(\{[\gamma_0]\}\big)>0$ and $\dim{(\vek H_{[\gamma_0]})}=1$ and otherwise $\dim{(\vek H_{[\gamma]})}=0$. Thus we may identify $\vek H$ with $\vek H_{[\gamma_0]}$. We immediately see that this means $\Lambda\subset[\gamma_0]$.

It follows, that for any rank-1-observable $M$ there is a weakly measurable unit-vector-valued function $\xi:\Lambda\to\mc M$ defining a decomposable isometry $W$ such that $(W\f)(\gamma)=\f(\gamma)\xi(\gamma)$ for all $\f\in\hil=L^2\big(\Lambda,\,\tilde\nu|_{\mc B(\Lambda)}\big)$ and a.a.\ $\gamma\in\Lambda$ which in turn defines the rank-1-observable. The integral operator $\mf W:\mc D\to\vek H=\vek H_{[\gamma_0]}\subset\mc M$ giving the minimal Kolmogorov decomposition for $M$ is of the form
$$
\mf W\f=\int_{H^\perp}\f(\gamma_0+\eta)\xi(\gamma_0+\eta)\,d\eta,\qquad\f\in\mc D,
$$
and thus it is clear that $\xi(\gamma)=c(\gamma)\xi_0$ with a measurable function $c:\Lambda\to\nT$ and a fixed $\xi_0\in\mc M$ so that $\dim{\vek H}=1$. Thus the covariance structure $\mc O_U(\Om,\,\hil)$ allows rank-1-observables if and only if there is a $[\gamma_0]\in\hat G/H^\perp$ such that $\nu\big(\{[\gamma_0]\}\big)>0$, $\Lambda\subset[\gamma_0]$, and $\dim{(\hil_\gamma)}=1$ for a.a.\ $\gamma\in\Lambda$. This result parallels \cite[Proposition 4]{car}.

One sees immediately that $M\in\mr{Ext}(\mc O_U(\Om,\,\hil))$ when $M$ is of rank 1 but it may happen that $M\notin\mr{Ext}(\mc O(\Om,\,\hil))$. For example, let $\hil=\nC^2$ with an orthonormal basis $\{|0\ra,\,|1\ra\}$, $U:\,\nT\to\mathcal L(\nC^2),\,z\mapsto U(z)=|0\ra\la 0|+z|1\ra\la 1|$, and $M:\,\mathcal B(\nT)\to\mathcal L(\nC^2),\;B\mapsto M(B)=\mu(B)(|0\ra\la 0|+|1\ra\la 1|)+\int_B\overline z \, d\mu(z)|0\ra\la 1|+\int_B z \, d\mu(z)|1\ra\la 0|$ a $(\nT,U)$-covariant POVM of rank 1 where $\mu$ is the normalized Haar measure on $\nT$. Hence, $M\in\mr{Ext}(\mc O_U(\nT,\,\nC^2))$ but $\int_\nT z^2 dM(z)=0$ so that $M\notin\mr{Ext}(\mc O(\nT,\,\nC^2))$.
\end{rem}

\begin{rem}\rm \label{remarkki}
In the case of the trivial subgroup $H=\{e\}$, where $e$ is the unit element of $G$, the result of Proposition \ref{prop1} can be simplified: The covariance structure $\mc O_U(G,\,\hil)$ is non-empty if and only if the measure $\nu_U$ is absolutely continuous with respect to the Haar measure $d\gamma$ of the dual group $\hat G$;
this result has also been obtained in \cite{hoa}. 
Indeed, now $G/H\simeq G$, $d\mu(g)=a\,dg$, $a>0$, (where $dg$ is some Haar measure)
$H^\perp=\hat G$, and $\hat G/H^\perp$ and $\nu$ are trivial so that $d\tilde\nu(\gamma)=b\,d\gamma$, $b>0$. The constants $a$ and $b$ are chosen such a way that $\mc F:L^2(G,a\,dg)\to L^2(\hat G,b\,d\gamma)$ is unitary, i.e.\ for all $f\in(L^1\cap L^2)(G,a\,dg)$, the Plancherel's formula
$$
\int_G\overline{f(g)}f(g)dg
=a^{-1}b\int_{\hat G}\overline{(\mc Ff)(\gamma)}(\mc Ff)(\gamma)d\gamma=
a b\int_{\hat G}\int_\Om\int_\Om\la g^{-1}g',\gamma\ra \overline{f(g)}f(g')\,dgdg'd\gamma
$$
holds. For example, when $G=\nR^n$, $\hat G=\nR^n$, and $dg$ and $d\gamma$ are Lebesgue measures, $ab=(2\pi)^{-n}$.
 
When $H=\{e\}$, $\hil$ can be chosen to be 
$\suor{\Lambda}\hil_\gamma\,d\gamma$ where $\Lambda\in\mathcal B(\hat G)$ is such that $\dim\hil_\gamma>0$ for all $\gamma\in\Lambda$, and
$
\mc D=\left\{\f\in\hil\,\big|\,\int_{\Lambda}\|\f(\gamma)\|\,d\gamma<\infty\right\}.
$
Let $W(\gamma):\,\hil_\gamma\to\mc M$ be the field of isometries associated to $M\in\mc O_U(G,\,\hil)$.
Now 
$\mf W:\mc D\to\mc M$ is defined by
$
\mf W\f=\int_{\Lambda} W(\gamma)\f(\gamma)\,d\gamma
$
for all $\f\in\mc D$ and $\vek H=\overline{\mf W(\mc D)}\subset\mc M$. 
Thus, $M\in\mr{Ext}(\mc O_U(G,\,\hil))$ if and only if $W(\gamma')^*AW(\gamma')=0$ for a.a.\ $\gamma'\in\Lambda$ implies $A=0$ (where $A\in\mathcal L(\vek H)$) \cite{ho3}.
Moreover, $M\in\mr{Ext}(\mc O(G,\,\hil))$ if and only if 
$$
\int_G\int_{\Lambda}\int_{\Lambda}\la\zeta-\xi,g\ra\sis{W(\zeta)\f(\zeta)}{D(g)W(\xi)\psi(\xi)}{}\,d\xi\,d\zeta\,d g=0
$$
for all $\f,\,\psi\in\mc D$,
implies $D(g)=0$ for a.a.\ $g\in G$ (where $D(g)\in\mathcal L(\vek H)$, $g\in G$, is an essentially bounded measurable family of operators).

In the following two sections we consider examples of the above theory where the value space $\Omega$ coincides with the symmetry group $G$, i.e.\ $H=\{e\}$. 
\end{rem}

\section{Covariant Position Observables}\label{cpo}

Suppose that $\Lambda\in\mc B(\nR^n)$ is not of Lebesgue measure zero. Consider a representation $U_\Lambda$ of $\nR^n$ in the Hilbert space $L^2(\Lambda)$ defined by
$$
\big(U_\Lambda(\vek q)\f\big)(\vek p)=e^{i(\vek q|\vek p)}\f(\vek p)
$$
for all $\vek q\in\nR^n$, $\vek p\in\Lambda$ and $\f\in L^2(\Lambda)$. Here $(\vek q|\vek p):=\sum_{k=1}^nq_kp_k$ for all $\vek q=(q_1,\ldots,q_n)\in\nR^n$ and $\vek p=(p_1,\ldots,p_n)\in\Lambda$. Thus, operators $U_\Lambda(\vek q)$ act diagonally. The measure $\nu_{U_\Lambda}$ associated with $U_\Lambda$ is defined by $\nu_{U_\Lambda}(B)=\ell(B\cap\Lambda)$ for any $B\in\mc B(\nR^n)$ where $\ell$ is the Lebesgue measure of $\nR^n$. Hence, $\mc O_{U_\Lambda}\big(\nR^n,\,L^2(\Lambda)\big)\neq\emptyset$. Fix an infinite-dimensional Hilbert space $\mc M$. According to Theorem \ref{theor1} any $M\in\mc O_{U_\Lambda}\big(\nR^n,\,L^2(\Lambda)\big)$ can be obtained by fixing a measurable field $\Lambda\ni\vek p\mapsto W(\vek p)\in\mc L(\nC;\,\mc M)$ of isometries. On the other hand, any such field can be fixed by picking any weakly measurable unit-vector-valued function $\xi:\Lambda\to\mc M$ and setting
$$
W(\vek p)\f(\vek p)=\f(\vek p)\xi(\vek p),\qquad\vek p\in\Lambda.
$$

The following proposition is a direct consequence of Theorems \ref{theor1}, \ref{theor2} and \ref{theor4}
(see Remark \ref{remarkki}).
\begin{prop}\label{extra}
Suppose that $M\in\mc O_{U_\Lambda}\big(\nR^n,\,L^2(\Lambda)\big)$. There is a weakly measurable unit-vector-valued function $\xi:\Lambda\to\mc M$ such that
\begin{equation}\label{lambdasuure}
\sis{\f}{M(B)\psi}{}=(2\pi)^{-n}\int_B\int_\Lambda\int_\Lambda e^{i(\vek x|\vek p_1-\vek p_2)}\sis{\xi(\vek p_1)}{\xi(\vek p_2)}{}\overline{\f(\vek p_1)}\psi(\vek p_2)\,d^n\vek p_1\,d^n\vek p_2\,d^n\vek x
\end{equation}
for all $\f,\,\psi\in(L^1\cap L^2)(\Lambda)$ and $B\in\mc B(\nR^n)$. On the other hand, given a measurable unit-vector-valued function $\xi:\Lambda\to\mc M$, \eqref{lambdasuure} defines an observable $M\in\mc O_{U_\Lambda}\big(\nR^n,\,L^2(\Lambda)\big)$. Suppose that $M$ is as in \eqref{lambdasuure}. Let $\vek H$ denote the Hilbert space generated by vectors $\int_{\Lambda}\f(\vek p)\xi(\vek p)\,d^n\vek p$, $\f\in(L^1\cap L^2)(\Lambda)$. The observable $M$ is extreme in $\mc O_{U_\Lambda}\big(\nR^n,\,L^2(\Lambda)\big)$ if and only if there is no nonzero $A\in\mc L(\vek H)$ such that
\begin{equation}\label{extlambda}
\sis{\xi(\vek p)}{A\xi(\vek p)}{}=0
\end{equation}
for a.a.\ $\vek p\in\Lambda$. The observable $M$ is extreme in $\mc O\big(\nR^n,\,L^2(\Lambda)\big)$ if and only if there is no non-zero decomposable operator $D\in\mc L\big(L^2(\nR^n;\,\vek H\big))$ with components $D(\vek x)\in\mc L(\vek H)$, $\vek x\in\nR^n$, such that
\begin{equation}
\int_{\nR^n}\int_{\nR^n}\int_{\nR^n}e^{i(\vek x|\vek p_1-\vek p_2)}\sis{\xi(\vek p_1)}{D(\vek x)\xi(\vek p_2)}{}\overline{\f(\vek p_1)}\psi(\vek p_2)\,d^n\vek p_1\,d^n\vek p_2\,d^n\vek x=0
\end{equation}
for all $\f,\,\psi\in (L^1\cap L^2)(\nR^n)$.
\end{prop}
It should be noted that we always find extreme observables in $\mc O_{U_\Lambda}\big(\nR^n,\,L^2(\Lambda)\big)$ which are not PVMs \cite{ho3}. Especially any observable as in \eqref{lambdasuure} is extreme in $\mc O_{U_\Lambda}\big(\nR^n,\,L^2(\Lambda)\big)$ if the unit-vector-valued function $\xi:\nR^n\to\mc M$ is such that the vectors $\xi(\vek p)$, $\vek p\in\nR^n$, generate a dense subspace of $\bf H$ \cite{ho3}; such functions clearly exist.

Consider the case $\Lambda=\nR^n$. We define the Fourier-Plancherel operator $\mc F:L^2(\nR^n)\to L^2(\nR^n)$ for all $\f\in(L^1\cap L^2)(\nR^n)$ by
$$
(\mc F\f)(\vek p)=(2\pi)^{-n/2}\int_{\nR^n}e^{i(\vek p|\vek x)}\f(\vek x)\,d^n\vek x,\qquad\vek p\in\nR^n.
$$
We often write $\mc F\f=\hat\f$ for all $\f\in L^2(\nR^n)$. Define a representation $U:\nR^n\ni\vek q\mapsto\mc F^*U_{\nR^n}(\vek q)\mc F$ in $L^2(\nR^n)$; in other words $\big(U(\vek q)\f\big)(\vek x)=\f(\vek x-\vek q)$ for all $\vek q,\,\vek x\in\nR^n$ ja $\f\in L^2(\nR^n)$. We may characterize the covariance structure $\mc O_U\big(\nR^n,\,L^2(\nR^n)\big)$ easily using the above results concerning $U_{\nR^n}$. For any $M\in\mc O_U\big(\nR^n,\,L^2(\nR^n)\big)$ there exists a weakly measurable unit-vector-valued function $\xi:\nR^n\to\mc M$ such that
\begin{equation}\label{mfmur}
\sis{\f}{M(B)\psi}{}=(2\pi)^{-n}\int_B\int_{\nR^n}\int_{\nR^n}e^{i(\vek x|\vek p_1-\vek p_2)}\sis{\xi(\vek p_1)}{\xi(\vek p_2)}{}\overline{\hat\f(\vek p_1)}\hat\psi(\vek p_2)\,d^n\vek p_1\,d^n\vek p_2\,d^n\vek x
\end{equation}
for all $B\in\mc B(\nR^n)$ and $\f,\,\psi\in C_c(\nR^n)$. The Hilbert space $L^2(\nR^n)$ gives the quantum mechanical description for a non-relativistic spin-0 particle moving in the space $\nR^n$. The representation $U$ is the {\it position translation representation} and the observables covariant in position translations are called {\it position observables}. Position observables are exhaustively characterized by \eqref{mfmur}. The results of Proposition \ref{extra} are also valid for position observables.

We may impose an additional requirement for a position observable: invariance under momentum boosts. This means that we call $M\in\mc O_U\big(\nR^n,\,L^2(\nR^n)\big)$ a {\it
(momentum boost) invariant position observable} if it satisfies the invariance condition
\begin{equation}\label{mominv}
V(\vek p)M(B)V(\vek p)^*=M(B)
\end{equation}
for all $\vek p\in\nR^n$ and $B\in\mc B(\nR^n)$. Here $V$ is the {\it momentum boost representation} of $\nR^n$ in $L^2(\nR^n)$, i.e.\ $\big(V(\vek p)\f\big)(\vek x)=e^{i(\vek p|\vek x)}\f(\vek x)$ for all $\vek p,\,\vek x\in\nR^n$ and $\f\in L^2(\nR^n)$. 
We denote the convex set of momentum boost invariant position observables by $\mf L(\nR^n)$. 

Suppose that $M\in\mc O_U\big(\nR^n,\,L^2(\nR^n)\big)$ is as in \eqref{mfmur}. 
Simple calculation shows that the condition \eqref{mominv} for $M$ is equivalent with
\begin{equation}\label{mom}
\sis{\xi(\vek p_1+\vek p)}{\xi(\vek p_2+\vek p)}{}=\sis{\xi(\vek p_1)}{\xi(\vek p_2)}{}
\end{equation}
for a.a.\ $\vek p_1,\,\vek p_2,\,\vek p\in\nR^n$. This means that we may replace the vector-valued function $\xi$ with a measurable function $\eta:\nR^n\to\nC$ such that the value $\eta(\vek p)$ coincides with the (essentially) constant value of the function
$$
\vek p_0\mapsto\sis{\xi(\vek p_0+\vek p)}{\xi(\vek p_0)}{}
$$
for a.a.\ $\vek p\in\nR^n$. We may now write for all $B\in\mc B(\nR^n)$ and $\f,\,\psi\in C_c(\nR^n)$
\begin{equation}\label{ps1}
\sis{\f}{M(B)\psi}{}=(2\pi)^{-n}\int_B\int_{\nR^n}\int_{\nR^n}e^{i(\vek x|\vek p_1-\vek p_2)}\eta(\vek p_1-\vek p_2)\overline{\hat\f(\vek p_1)}\hat\psi(\vek p_2)\,d^n\vek p_1\,d^n\vek p_2\,d^n\vek x.
\end{equation}

Since $\int_{\nR^n}\int_{\nR^n}\sis{\xi(\vek p_1)}{\xi(\vek p_2)}{}\overline{f(\vek p_1)}f(\vek p_2)\,d^n\vek p_1\,d^n\vek p_2\geq0$ for all $f\in C_c(\nR^n)$, we note that $\eta$ is a function of positive type, i.e.\
$$
\int_{\nR^n}\eta(\vek p)(f^**f)(\vek p)\,d^n\vek p\geq0
$$
for all $f\in L^1(\nR^n)$. Here $f^*(\vek p)=\overline{f(-\vek p)}$ for a.a.\ $\vek p\in\nR^n$ and the bilinear operator $*$ is the convolution in $L^1(\nR^n)\times L^1(\nR^n)$. Any function of positive type coincides almost everywhere with a single continuous function \cite{fol} and we may thus assume that $\eta$ is continuous. The normalization condition $\|\xi(\vek p)\|=1$ for a.a.\ $\vek p\in\nR^n$ now reads $\eta(\veg 0)=1$. Let us denote the convex set of continuous functions $\eta:\nR^n\to\nC$ of positive type with $\eta(\veg0)=1$ by $\mf E(\nR^n)$. Any such function defines a positive sesquilinear form $S$ on $L^2(\nR^n)$ through
$$
S(f,g)=\int_{\nR^n}\int_{\nR^n}\eta(\vek p_1-\vek p_2)\overline{f(\vek p_1)}g(\vek p_2)\,d^n\vek p_1\,d^n\vek p_2\qquad f,\,g\in C_c(\nR^n).
$$
From this and the condition $\eta(\veg0)=1$ we obtain that there is a weakly measurable unit vector valued function $\xi:\nR^n\to L^2(\nR^n)$ with the property \eqref{mom} such that $\sis{\xi(\vek p_1)}{\xi(\vek p_2)}{}=\eta(\vek p_1-\vek p_2)$ for a.a.\ $\vek p_1,\,\vek p_2\in\nR^n$. It is now clear that there is an affine one-to-one correspondence between the functions $\eta\in\mf E(\nR^n)$ and invariant position observables $M$ given by \eqref{ps1}.

The Bochner theorem states that for every function of positive type $\zeta:\nR^n\to\nC$ there is a positive Borel measure $\mu$ on $\nR^n$ such that
$$
\zeta(\vek p)=\int_{\nR^n}e^{-i(\vek p|\vek x)}\,d\mu(\vek x),\qquad\vek p\in\nR^n.
$$
Especially for all $\eta\in\mf E(\nR^n)$ the corresponding measure $\rho$ is a probability measure. In addition any probability measure $\rho:\mc B(\nR^n)\to[0,1]$ defines a function $\eta\in\mf E(\nR^n)$ in this manner. Suppose that the invariant position observable $M$ is as in \eqref{ps1} with a function $\eta\in\mf E(\nR^n)$ arising from a probability measure $\rho$, i.e.
$$
\eta(\vek p)=\int_{\nR^n}e^{-i(\vek p|\vek x)}\,d\rho(\vek x),\qquad\vek p\in\nR^n.
$$
From this and the characterization \eqref{ps1} of position observables we obtain easily that $M=\rho*M_\ka$, where $M_\ka$ is the {\it canonical position observable,} $M_\ka(B)\f=\chi_B\f$ for all $B\in\mc B(\nR^n)$ and $\f\in L^2(\nR^n)$. The convolution $\rho*M_\ka$ is given by
$$
\sis{\f}{\rho*M_\ka(B)\psi}{}=\int_B\int_{\nR^n}\overline{\f(\vek x-\vek q)}\psi(\vek x-\vek q)\,d\rho(\vek q)\,d^n\vek x
$$
for all $B\in\mc B(\nR^n)$ and $\f,\,\psi\in L^2(\nR^n)$. It is known \cite[Theorems 3.20 and 3.25]{fol} that the extreme points of the convex set $\mf E(\nR^n)$ are the group homomorphisms $\vek p\mapsto e^{i(\vek p|\vek q)}$ for some $\vek q\in\nR^n$ corresponding to point measures $\delta_{\vek q}$ and translated canonical position observables $M_{\vek q}$ such that $M_{\vek q}(B)=M_\ka(B-\vek q)$, $B\in\mc B(\nR^n)$. We have thus obtained the following well known \cite{ca2} result:
\begin{prop}\label{extra0}
For any momentum boost invariant position observable $M\in\mf L(\nR^n)$, there exists a probability measure $\rho:\mc B(\nR^n)\to[0,1]$ such that $M=\rho*M_\ka$. Moreover, the correspondence between invariant position observables $M\in\mf L(\nR^n)$ and probability measures $\rho:\mc B(\nR^n)\to[0,1]$ is affine and bijective. The extreme points of $\mf L(\nR^n)$ are all PVMs and of the form $M_{\vek q}$, $M_{\vek q}(B)=M_\ka(B-\vek q)$, $B\in\mc B(\nR^n)$, for some $\vek q\in\nR^n$.
\end{prop}

We close this section with a related example. Consider the Hilbert space $L^2(\nT^n,\,\mu_n)$, where
$$
d\mu_n(\vek z)=(2\pi)^{-n}d(\arg{(z_1)})\cdots d(\arg{(z_2)}),\qquad\vek z=(z_1,\ldots,z_n)\in\nT^n.
$$
This space has an orthonormal basis $\{e_{\vek m}\,|\,\vek m\in\nZ^n\}$ such that $e_{\vek m}(\vek w)=\overline{\la\vek w,\vek m\ra}$ for all $\vek m\in\nZ^n$ and $\vek w\in\nT^n$. The dual action of $\nZ^n$ on $\nT^n$ is defined by
$$
\la\vek w,\vek m\ra=w_1^{m_1}\cdots w_n^{m_n},\quad\vek w=(w_1,\ldots,w_n)\in\nT^n,\quad\vek m=(m_1,\ldots,m_n)\in\nZ^n.
$$
Consider now a Hilbert space $\hil_Z$ spanned by an orthonormal basis $\{e_{\vek m}\,|\,\vek m\in Z\}$ with $Z\subset\nZ^n$. Define a unitary representation $U_Z$ of $\nT^n$ in $\hil_Z$ through
$$
U_Z(\vek w)=\sum_{\vek m\in Z}\la\vek w,\vek m\ra|e_{\vek m}\ra\la e_{\vek m}|,\qquad\vek w\in\nT^n.
$$
The case $Z=\nZ^n$ corresponds to {\it position observables of a particle confined to the compact cyclic space $\nT^n$.} The representation $U_{\nZ^n}=:U$ acts now as
$$
\big(U(\vek w)\f\big)(\vek z)=\f(\vek z\overline{\vek w})
$$
for all $\vek w,\,\vek z\in\nT^n$ and $\f\in\hil_{\nZ^n}=L^2(\nT^n,\mu_n)$. Here $\overline{\vek w}=\overline{(w_1,\ldots,w_n)}=(\overline{w_1},\ldots,\overline{w_n})$.

The covariance structure 
$\mc O_{U_Z}\big(\nT^n,\,\hil_Z\big)$ and its extreme points can easily be characterized as in the beginning of the section. We give the results without proofs.
\begin{prop}\label{extra1}
For any $M\in\mc O_{U_Z}\big(\nT^n,\,\hil_Z\big)$ there is a family $\{\xi_{\vek m}\,|\,\vek m\in Z\}$ of unit vectors in an infinite dimensional Hilbert space $\mc M$ such that
\begin{equation}\label{vaih}
M(B)=\sum_{\vek k,\,\vek l\in Z}\sis{\xi_{\vek k}}{\xi_{\vek l}}{}\int_B\la\vek z,\vek k-\vek l\ra\,d\mu_n(\vek z)\,|e_{\vek k}\ra\la e_{\vek l}|,\qquad B\in\mc B(\nT^n).
\end{equation}
On the other hand, an observable $M$ as in \eqref{vaih} is $U_Z$-covariant with any choice of the unit vectors $\xi_{\vek m}\in\mc M$, $\vek m\in Z$. Suppose that $M$ is a covariant observable as in \eqref{vaih}. Denote the Hilbert space generated by the vectors $\xi_{\vek m}$, $\vek m\in Z$, by $\vek H$. The observable $M$ is extreme in $\mf M^{U_Z}_{\nT^n}$ if and only if there is no non-zero operator $A\in\mc L(\vek H)$ such that
\begin{equation}
\sis{\xi_{\vek m}}{A\xi_{\vek m}}{}=0
\end{equation}
for all $\vek m\in Z$. The observable $M$ is extreme in the set $\mc O(\nT^n,\,\hil)$ if and only if there is no  non-zero decomposable operator $D\in\mc L\big(L^2(\nT^n,\mu_n;\vek H)\big)$ with components $D(\vek z)\in\mc L(\vek H)$ such that
\begin{equation}
\int_{\nT^n}\la\vek z,\vek k-\vek l\ra\sis{\xi_{\vek k}}{D(\vek z)\xi_{\vek l}}{}\,d\mu_n(\vek z)=0
\end{equation}
for all $\vek k,\,\vek l\in Z$.
\end{prop}

In the case $Z=\nZ^n$, similarly as before, one can define the {\it canonical position observable} of a particle moving on $\nT^n$. It is even easier to show that all momentum shift invariant position observables are convolutions of probability measures on $\nT^n$ with the canonical position. The corresponding extreme observables are shifted canonical position observables and, hence, they are PVMs.

One may easily check that if $Z\neq\nZ^n$ the covariance structure $\mc O_{U_Z}\big(\nT^n,\,\hil_Z\big)$ contains no PVMs. In the case $n=1$ the choice $Z=\{0,\,1,\,2,\ldots\}$ corresponds to {\it covariant phase observables} of an electromagnetic mode. Thus, a phase observable is never a PVM. Choosing the vectors $\xi_k$, $k=0,\,1,\,2,\ldots$, of \eqref{vaih} such that $\xi_k=\xi_0$ for all $k=0,\,1,\,2,\ldots$ we obtain the {\it canonical phase observable} $\F$. One may easily check by using Proposition \ref{extra1} that $\F$ is extreme in $\mc O(\nT,\,\hil)$. This result was obtained already in \cite{he2}.

\section{Time Observables of a Free Particle}\label{tofp}

We investigate another example where $H=\{e\}$. Consider a free non-relativistic spin-0 particle of mass $m$ moving on a line. The Hilbert space of the system is $L^2(\nR)$. Let $P$ be the usual momentum operator, i.e.\ the extension of the differential operator $P\f=-i\f'$ defined densely in $L^2(\nR)$. The generator of time shifts is the Hamiltonian $H_0=P^2/(2m)$ and it defines a representation $V$ of the group of time shifts (group of additive real numbers) by $V(t)=e^{itH_0}$ for all $t\in\nR$. Since the spectrum of $H_0$ is bounded from below\footnote{Indeed, $\sigma(H_0)=[0,\infty)$.} there is no self-adjoint operator $T$ on $L^2(\nR)$ canonically conjugated with $H_0$, i.e.
\begin{equation}\label{aikakomm}
[T,H_0]\f=TH_0\f-H_0T\f=-i\f
\end{equation}
for all $\f$ in some dense subspace of $L^2(\nR)$. This in turn means that there is no PVM $P\in\Sigma\big(\nR,\,L^2(\nR)\big)$ with the covariance property
\begin{equation}\label{aika}
V(t)P(B)V(t)^*=P(B+t)
\end{equation}
for all $t\in\nR$ and $B\in\mc B(\nR)$. However, there are POVMs $M\in\mc O\big(\nR,\,L^2(\nR)\big)$ covariant with respect to $V$. We call these observables $M\in\mc O_V\big(\nR,\,L^2(\nR)\big)$ \emph{(covariant) time observables of a free particle}. For every $\f\in L^2(\nR)$ denote its Fourier-Plancherel transform by $\hat\f$ and especially if $\f\in(L^1\cap L^2)(\nR)$
$$
\hat\f(p)=\frac{1}{\sqrt{2\pi}}\int_\nR e^{ipx}\f(x)\,dx,\qquad p\in\nR.
$$

Let us denote the set of positive real numbers by $\nR^+$. Define a unitary map $L^2(\nR)\ni\f\mapsto\tilde\f\in L^2(\nR^+;\,\nC^2)$ as in \cite[III.8]{hoi}, where
$$
\tilde\f(\e)=\left(\frac{m}{2\e}\right)^{1/4}\big(\hat\f(\sqrt{2m\e}),\hat\f(-\sqrt{2m\e})\big),\qquad\e>0,
$$
for all $\f\in L^2(\nR)$. One has
$$
\big(\widetilde{V(t)\f}\big)(\e)=e^{i\e t}\tilde\f(\e)
$$
for all $\f\in L^2(\nR)$, $t\in\nR$ and $\e>0$ \cite{hoi}. We define the dual action of $\nR$ onto itself by $\la t,\e\ra=e^{i\e t}$. The component spaces in the direct-integral representation \eqref{suora} are thus $\nC^2$. The value of a vector $\f\in L^2(\nR)$ in the fiber defined by an $\e>0$ is the vector $\tilde\f(\e)\in\nC^2$. The measure $\nu_V:\mc B(\nR)\to[0,\infty]$ in \eqref{suora} is simply the $\nR^+$-supported Lebesgue measure. Hence there are time observables covariant with respect to $V$.

Fix an infinite-dimensional Hilbert space $\mc M$ and a weakly measurable field $\nR^+\ni\e\mapsto W(\e)\in\mc L(\nC^2;\,\mc M)$ of isometries. Using Theorem \ref{theor1} we see that the operator valued set function $M:\mc B(\nR)\to\mc L\big(L^2(\nR)\big)$ defined by
\begin{equation}\label{aiak}
\sis{\f}{M(B)\psi}{}=\frac{1}{2\pi}\int_B\int_0^\infty\int_0^\infty e^{it(\e_2-\e_1)}\sis{\tilde\f(\e_1)}{W(\e_1)^*W(\e_2)\tilde\psi(\e_2)}{}\,d\e_1\,d\e_2\,dt
\end{equation}
for all $B\in\mc B(\nR)$ and $\f,\,\psi\in S(\nR)$ is a covariant time observable. Here $S(\nR)$ is the Schwartz space of rapidly decreasing functions $\nR\to\nC$. On the other hand, given a time observable $M\in\mc O_V\big(\nR,\,L^2(\nR)\big)$, there is always a field $\e\mapsto W(\e)$ of isometries that defines $M$ through \eqref{aiak}. We may determine any such field of isometries by fixing unit-vector-valued weakly measurable functions $\zeta_j:\nR^+\to\mc M$, $j=0,\,1$, such that $\zeta_0(p)\perp\zeta_1(p)$ for a.a.\ $p>0$ and setting
$$
W(\e)=|\zeta_0(\sqrt{2m\e})\ra\la0|+|\zeta_1(\sqrt{2m\e})\ra\la1|,\qquad\e>0,
$$
where the vectors $|0\ra$ and $|1\ra$ constitute an orthonormal basis of $\nC^2$. We obtain the following result as a direct consequence of the Theorems \ref{theor1}, \ref{theor2} and \ref{theor4}.

\begin{prop}\label{prop7}
For any time observable $M\in\mc O_V\big(\nR,\,L^2(\nR)\big)$ there exist unit-vector-valued weakly measurable functions $\zeta_j:\nR^+\to\mc M$, $j=0,\,1$, with $\zeta_0(p)\perp\zeta_1(p)$ for a.a.\ $p>0$ such that
\begin{equation}\label{aikasuure}
\sis{\f}{M(B)\psi}{}=\frac{1}{2\pi m}\int_B\int_0^\infty\int_0^\infty e^{\frac{it}{2m}(p_2^2-p_1^2)}\sum_{j,\,k=0}^1\sis{\zeta_j(p_1)}{\zeta_k(p_2)}{}\overline{\hat\f\big((-1)^jp_1\big)}\hat\psi\big((-1)^kp_2\big)\sqrt{p_1p_2}\,dp_1\,dp_2\,dt
\end{equation}
for all $B\in\mc B(\nR)$ and $\f,\,\psi\in S(\nR)$. On the other hand, if the weakly measurable unit-vector valued functions $\zeta_j:\nR^+\to\mc M$ are as above, \eqref{aikasuure} defines a time observable. Suppose that a time observable $M$ is as in \eqref{aikasuure}. Denote by $\vek H$ the Hilbert space generated by the vectors $\int_0^\infty\big(\hat\f(p)\zeta_0(p)+\hat\f(-p)\zeta_1(p)\big)\sqrt{p}\,dp$ where $\f\in S(\nR)$. The observable $M$ is extreme in $\mc O_V\big(\nR,\,L^2(\nR)\big)$ if and only if there is no non-zero operator $A\in\mc L(\vek H)$ such that
\begin{equation}
\sis{\zeta_j(p)}{A\zeta_k(p)}{}=0
\end{equation}
for all $j,\,k=0,\,1$ and a.a.\ $p>0$. The observable $M$ defined as in \eqref{aikasuure} is extreme in $\mc O\big(\nR,\,L^2(\nR)\big)$ if and only if there is no non-zero decomposable operator $D\in\mc L\big(L^2(\nR;\vek H)\big)$ with components $D(t)\in\mc L(\vek H)$, $ t\in\nR$, such that
$$
\int_{\nR}\int_0^\infty\int_0^\infty e^{\frac{it}{2m}(p_2^2-p_1^2)}\sum_{j,\,k=0}^1\sis{\zeta_j(p_1)}{D(t)\zeta_k(p_2)}{}\overline{\hat\f\big((-1)^jp_1\big)}\hat\psi\big((-1)^kp_2\big)\sqrt{p_1p_2}\,dp_1\,dp_2\,dt=0
$$
for all $\f,\,\psi\in S(\nR)$.
\end{prop}

Let us take a closer look at the special time observable $\tau\in\mc O_V\big(\nR,\,L^2(\nR)\big)$ which is defined by \eqref{aikasuure} with constant functions $\zeta_j(p)=\f_j$ for $j=0,\,1$ and a.a.\ $p\geq0$ where $\f_0\perp\f_1$. Thus for any $\f,\,\psi\in S(\nR)$ and all $B\in\mc B(\nR)$
\begin{equation}
\sis{\f}{\tau(B)\psi}{}=\frac{1}{2\pi m}\int_B\int_0^\infty\int_0^\infty e^{\frac{it}{2m}(p_2^2-p_1^2)}\big(\overline{\hat\f(p_1)}\hat\psi(p_2)+\overline{\hat\f(-p_1)}\hat\psi(-p_2)\big)\sqrt{p_1p_2}\,dp_1\,dp_2\,dt.
\end{equation}
We call $\tau$ the \emph{canonical time observable (of a free particle)} \cite{hoi}. It is clear that $\tau$ is an extreme observable in the set $\mc O_V\big(\nR,\,L^2(\nR)\big)$. Next we show that it is also extreme in $\mc O\big(\nR,\,L^2(\nR)\big)$. 

Suppose that there is a bounded decomposable operator $D\in\mc L\big(L^2(\nR;\vek H)\big)$ with components $D(t)\in\mc L(\vek H)$, $t\in\nR$, where $\vek H$ is just the two-dimensional space generated by the vectors $\f_0$ and $\f_1$ such that
$$
\int_\nR\int_0^\infty\int_0^\infty e^{\frac{it}{2m}(p_2^2-p_1^2)}\sum_{j,\,k=0}^1\overline{\hat\f\big((-1)^jp_1\big)}\hat\psi\big((-1)^kp_2\big)D_{j,k}(t)\sqrt{p_1p_2}\,dp_1\,dp_2\,dt=0
$$
for all $k,\,l=0,\,1,\,2,\ldots$ and $\f,\,\psi\in S(\nR)$ where $D_{j,k}(t):=\sis{\f_j}{D(t)\f_k}{}$ for all $t\in\nR$ and $j,\,k=0,\,1$. Assume, for example, that in the above formula $\f$ is such that $\hat\f$ is supported by $\nR^+$ and that $\psi$ is such that $\hat\psi$ is supported by the complement of $\nR^+$. Now only the term $j=0,\,k=1$ of the sum is left in the above formula. Proceeding in a similar fashion we can isolate all the terms of the sum and we conclude that
$$
\int_\nR\int_0^\infty\int_0^\infty e^{it(\e_1-\e_2)}g(\e_1)f(\e_2)D_{j,k}(t)\,d\e_1\,d\e_2\,dt=0
$$
for all $f,\,g\in (L^1\cap L^2)(\nR)$ supported by $\nR^+$ and $j,\,k=0,\,1$. Define the parity operator $\mc P$, $(\mc P\f)(x)=\f(-x)$ for all $\f\in L^2(\nR)$ and a.a.\ $x\in\nR$. We may rewrite the above equation using the properties of the Fourier transform in the form
$$
\int_\nR(\widehat{f*\mc Pg})(t)D_{j,k}(t)\,dt=0,
$$
where $*$ is the convolution. To show that $D=0$ it thus suffices to show that the linear space spanned by vectors $\widehat{f*\mc Pg}$, where $f,\,g\in(L^1\cap L^2)(\nR)$ are supported by $\nR^+$, is dense in $L^1(\nR)$. We define (associated) Laguerre polynomials $L_n^j$, $n\in\nN$, $j\in\nR$, through
$$
L_n^j(x)=\frac{x^{-j}e^x}{n!}\frac{d^n}{dx^n}\big(x^{n+j}e^{-x}\big),\qquad x>0.
$$
We denote $L_n^0=:L_n$ for all $n\in\nN$. Consider the scaled Laguerre polynomials $f(x)=L_m(x)e^{-x/2}$ and $g(x)=L_n(x)e^{-x/2}$ for all $x>0$. Using formulas
$$
L_n(x+y)=\sum_{k=0}^nL_k(x)L_{n-k}^{-1}(y),\qquad\int_0^\infty L_m(x)L_n(x)e^{-x}\,dx=\delta_{m,n}
$$
one obtains
$$
(f*\mc Pg)(u)=\left\{\begin{array}{ll}L_{n-m}^{-1}(-u)e^{u/2},&u\leq0\\0,&u>0\end{array}\right.
$$
when $m<n$,
$$
(f*\mc Pg)(u)=\left\{\begin{array}{ll}0,&u\leq0\\L_{m-n}^{-1}(u)e^{-u/2},&u>0\end{array}\right.
$$
when $n<m$ and
$$
(f*\mc Pg)(u)=e^{-|u|/2}
$$
when $m=n$. Especially, these vectors generate functions of the form $\nR\ni x\mapsto x^ne^{-|x|}$ and thus their Fourier transforms span a dense subspace of $L^1(\nR)$. 
Hence, $D=0$ and thus $\tau$ is extreme in $\mc O\big(\nR,\,L^2(\nR)\big)$.

It is known \cite{hoi,bu3} that the first moment operator
$$
T:=\tau[1]=\int_\nR t\,d\tau(t)
$$
is symmetric and densely defined and that it coincides with the symmetric operator
$$
T'=m\,\mr{sign}(P)|P|^{-1/2}Q|P|^{-1/2}
$$
on a dense subspace of $L^2(\nR)$. Here $Q$ is the usual position operator: $(Q\f)(x)=x\f(x)$. Neither $T$ nor $T'$ is self-adjoint but $T'$ satisfies formally the commutation relation \eqref{aikakomm}.

\begin{rem}\rm
We may also define the time observables using Hermite functions $h_k\in L^2(\nR)$, $k=0,\,1,\,2,\ldots$, where
$$
h_k(x)=(-1)^k(2^kk!\sqrt\pi)^{-1/2}e^{x^2/2}\frac{d^k}{dx^k}e^{-x^2},\qquad x\in\nR,
$$
for all $k=0,\,1,\,2,\ldots$. Suppose that $k\in\nZ$. Define $\overline k=0$ if $2\mid k$ and $\overline k=1$ otherwise. The Hermite functions form an orthonormal basis of $L^2(\nR)$ and $\tilde h_k(\e)=i^k(2m/\e)^{1/4}h_k(\sqrt{2m\e})e_{\overline k}$ for all $k=0,\,1,\,2,\ldots$ and $\e>0$. Here
$$
e_j=\frac{1}{\sqrt2}\big(1,(-1)^j\big)\in\nC^2,\qquad j=0,\,1.
$$
Using the Hermite basis and Proposition \ref{prop7} one finds that any time observable $M$ can be given in the form
$$
M(B)=\sum_{k,\,l=0}^\infty\frac{i^{l-k}}{\pi m}\int_B\int_0^\infty\int_0^\infty e^{\frac{it}{2m}(p_2^2-p_1^2)}h_k(p_1)h_l(p_2)\sis{\zeta_{\overline k}(p_1)}{\zeta_{\overline l}(p_2)}{}\sqrt{p_1p_2}\,dp_1\,dp_2\,dt\,|h_k\ra\la h_l|
$$
weakly for all $B\in\mc B(\nR)$ with some unit-vector-valued functions $\zeta_j:\nR^+\to\mc M$, $j=0,\,1$, with $\zeta_0(p)\perp\zeta_1(p)$ for a.a.\ $p>0$. 
\end{rem}

\section{Covariant Position Difference Observables}\label{cpdo}

Finally, we examine a situation where the transitive space $\Om$ differs from the symmetry group; covariant position difference observables. Our physical system consists of two (nonrelativistic) particles moving in the space $\nR^n$. The physical variable measured is the difference $\vek x=\vek x_2-\vek x_1$ of the positions $\vek x_1,\,\vek x_2\in\nR^n$ of the particles 1 and 2 respectively. When the positions of the particles are translated with a vector $\vek g=(\vek u_1,\vek u_2)\in\nR^{2n}$, $\vek u_1,\,\vek u_2\in\nR^n$ the position difference $\vek x\in\nR^n$ changes according to
\begin{equation}\label{nrtr}
\vek g\cdot\vek x=
(\vek u_1,\vek u_2)\cdot\vek x=\vek x+\vek u_2-\vek u_1.
\end{equation}
The position difference value space $\nR^n$ is homeomorphic to the coset space $\nR^{2n}/H$ where
$$
H=\{(\vek u,\vek u)\,|\,\vek u\in\nR^n\}
$$
is a closed subgroup of the additive translation group $\nR^{2n}$. The character group $\widehat{\nR^{2n}}$ is homeomorphic to $\nR^{2n}$ when the duality is given by $\la\vek g,\vek v\ra=e^{i(\vek g|\vek v)}$ for all $\vek g,\,\vek v\in\nR^{2n}$ and the annihilator of $H$ is
$$
H^\perp=\{(-\vek p,\vek p)\,|\,\vek p\in\nR^n\}.
$$
We identify the coset space $\nR^{2n}/H^\perp$ with $\nR^n$ by picking unique representatives $(\vek w,\veg0)$, $\vek w\in\nR^n$, from each coset.

The Hilbert space for the system is $L^2(\nR^{2n},d^{2n}\vek g)$. The first $n$ coordinates are assigned to the particle 1 and the remaining coordinates to the particle 2. Let us define a strongly continuous unitary representation $U$ of $\nR^{2n}$ in the space $L^2(\nR^{2n})$ by
$$
\big(U(\vek g)\f\big)(\vek g')=\f(\vek g'-\vek g)
$$
for all $\vek g,\,\vek g'\in\nR^{2n}$ and $\f\in L^2(\nR^{2n})$. One may check that $\mc FU(\vek g)=V(\vek g)\mc F$ where $\big(V(\vek g)\f\big)(\vek g')=e^{i(\vek g|\vek g')}\f(\vek g')$ for all $\vek g,\,\vek g'\in\nR^{2n}$ and $\f\in L^2(\nR^{2n})$. We may characterize the position difference observables in $\hil$ as the covariance structure $\mc O_U\big(\nR^n,\,L^2(\nR^{2n})\big)$, i.e.\ a {\it position difference observable} $M$ satisfies the covariance condition
$$
U(\vek g)M(B)U(\vek g)^*=M(\vek g\cdot B)
$$
for all $\vek g\in\nR^{2n}$ and $B\in\mc B(\nR^n)$ where the group $\nR^{2n}$ operates in the space $\nR^n$ according to \eqref{nrtr}. 

Choose the scaled Lebesgue-measure $d^n\vek p/(2\pi)^{n/2}$ as the Haar measure for the annihilator and also as the $\nR^{2n}$-invariant measure for the position difference value space, i.e.\ $d\mu(\vek x^n)=d^n\vek x/(2\pi)^{n/2}$. The measure $\nu$ on $\nR^{2n}/H^\perp\simeq\nR^n$ is chosen to be 
$d^n\vek w/(2\pi)^{n/2}$ so that
$\nu_U=\tilde\nu$ is $d^{2n}\vek v/(2\pi)^{n}$. Thus $\mc O_U\big(\nR^n,\,L^2(\nR^{2n})\big)\neq\emptyset$.

Since $L^2(\nR^{2n})$ is infinite-dimensional we may characterize all members $M$ of the covariance structure $\mc O_U\big(\nR^n,\,L^2(\nR^{2n})\big)$ by fixing a weakly measurable unit-vector-valued map $\xi:\nR^{2n}\to L^2(\nR^{2n})$ and defining the decomposable isometry $W$ of Theorem \ref{theor1} by $(W\hat\f)(\vek w)=\hat\f(\vek w)\xi(\vek w)$ for all $\f\in L^2(\nR^{2n})$. For each such isometry we may also define a measurable function $\alpha:\nR^{3n}\to\nC$ by
\begin{equation}\label{alf}
\alpha(\vek w,\vek p_1,\vek p_2)=\sis{\xi(\vek w-\vek p_1,\vek p_1)}{\xi(\vek w-\vek p_2,\vek p_2)}{}\qquad\vek w,\,\vek p_1,\,\vek p_2\in\nR^n.
\end{equation}
Let us now define the set $\mf A(\nR^n)$ of measurable functions $\alpha:\nR^{3n}\to\nC$ such that
\begin{enumerate}
\item $\alpha(\vek w,\vek p_1,\vek p_2)=\overline{\alpha(\vek w,\vek p_2,\vek p_1)}$ for a.a.\ $\vek w,\,\vek p_1,\,\vek p_2\in\nR^n$,
\item $\int_{\nR^n}\int_{\nR^n}\alpha(\vek w,\vek p_1,\,\vek p_2)\overline{f(\vek p_1)}f(\vek p_2)\,d^n\vek p_1\,d^n\vek p_2\geq0$ for all $f\in C_c(\nR^n)$ and for a.a.\ $\vek w\in\nR^n$ and
\item $\alpha^{(0)}(\vek w,\vek p)=1$ for a.a.\ $\vek w,\,\vek p\in\nR^n$.
\end{enumerate}
The number $\alpha^{(0)}(\vek w,\vek p)=\sum_{k\in\nN}|\alpha_k(\vek w,\vek p)|^2$ is the `diagonal value' given by the Kolmogorov decomposition
\begin{equation}\label{alhaj}
\alpha(\vek w,\vek p_1,\vek p_2)=\sum_{k\in\nN}\overline{\alpha_k(\vek w,\vek p_1)}\alpha_k(\vek w,\vek p_2)
\end{equation}
where the functions $\alpha_k:\nR^{2n}\to\nC$ are measurable. Such a decomposition can always be constructed since any $\alpha\in\mf A(\nR^n)$ gives kernels $(\vek p_1,\vek p_2)\mapsto\alpha(\vek w,\vek p_1,\vek p_2)$, $\vek w\in\nR^n$, for positive sesquilinear forms $S_{\vek w}$ determined by
$$
S_{\vek w}(f,g)=\int_{\nR^n}\int_{\nR^n}\alpha(\vek w,\vek p_1,\vek p_2)\overline{f(\vek p_1)}g(\vek p_2)\,d^n\vek p_1\,d^n\vek p_2
$$
for all $f,\,g\in C_c(\nR^n)$. The condition 3 ensures that by setting $\xi(\vek p_1,\vek p_2)=\sum_{k\in\nN}\alpha_k(\vek p_1,\vek p_2)e_k$, $\vek p_1,\,\vek p_2\in\nR^n$, with an orthonormal basis $\{e_k\,|\,k\in\nN\}\subset L^2(\nR^{2n})$ any $\alpha\in\mf A(\nR^n)$ with a decomposition as in \eqref{alhaj} determines a weakly measurable unit-vector-valued map $\xi:\nR^{2n}\to L^2(\nR^{2n})$. We may now characterize the covariance structure $\mc O_U\big(\nR^n,\,L^2(\nR^{2n})\big)$ and its extreme points.

\begin{prop}\label{prop6}
For any $M\in\mc O_U\big(\nR^n,\,L^2(\nR^{2n})\big)$ there is a function $\alpha\in\mf A(\nR^n)$ and a weakly measurable unit-vector-valued map $\xi:\nR^{2n}\to L^2(\nR^{2n})$ related to $\alpha$ according to \eqref{alf} such that
\begin{eqnarray}\label{prae}
\sis{\f}{M(B)\psi}{}&=&\frac1{(2\pi)^n}\int_B\int_{\nR^n}\int_{\nR^n}\int_{\nR^n}e^{i(\vek x|\vek p_1-\vek p_2)}\alpha(\vek w,\vek p_1,\vek p_2)\overline{\hat\f(\vek w-\vek p_1,\vek p_1)}\hat\psi(\vek w-\vek p_2,\vek p_2)\times\nonumber\\
&\times&d^n\vek p_1\,d^n\vek p_2\,d^n\vek w\,d^n\vek x
\end{eqnarray}
for all $\f,\,\psi\in C_c(\nR^{2n})$ and $B\in\mc B(\nR^n)$. On the other hand, given a function $\alpha\in\mf A(\nR^n)$, \eqref{prae} determines an observable $M\in\mc O_U\big(\nR^n,\,L^2(\nR^{2n})\big)$. Futhermore \eqref{prae} sets an affine bijection between $\mc O_U\big(\nR^n,\,L^2(\nR^{2n})\big)$ and $\mf A(\nR^n)$.

Suppose that $M\in\mc O_U\big(\nR^n,\,L^2(\nR^{2n})\big)$ is of the form \eqref{prae} with a unit-vector-valued map $\xi:\nR^{2n}\to L^2(\nR^{2n})$ as in \eqref{alf}. Let us denote the space generated by the vectors $\int_{\nR^n}\hat\f(\vek w-\vek p,\vek p)\xi(\vek w-\vek p,\vek p)\,d^n\vek p$ where $\f\in C_c(\nR^{2n})$ by $\vek H_{\vek w}$ for all $\vek w\in\nR^n$ and define the direct integral space $\vek H=\suor{\nR^n}\vek H_{\vek w}\,d^n\vek w$. The observable $M$ is an extreme covariant position difference observable, $M\in\mr{Ext}\Big(\mc O_U\big(\nR^n,\,L^2(\nR^{2n})\big)\Big)$, if and only if there is no non-zero decomposable operator $A\in\mc L(\vek H)$ with components $A(\vek w)\in\vek H_{\vek w}$, $\vek w\in\nR^n$, such that
\begin{equation}\label{extehto1}
\sis{\xi(\vek w-\vek p,\vek p)}{A(\vek w)\xi(\vek w-\vek p,\vek p)}{}=0
\end{equation}
for a.a.\ $\vek w,\,\vek p\in\nR^n$. The observable $M$ is extreme in $\mc O(\nR^n,\,\hil)$ if and only if there is no non-zero decomposable operator $D\in\mc L\big(L^2(\nR^n;\vek H)\big)$ with components $D(\vek w)\in\mc L(\vek H)$, $\vek w\in\nR^n$, such that
\begin{eqnarray}\label{extehto3}
&&\int_{\nR^n}e^{i(\vek x|\vek p_2-\vek p_1)}\sis{\xi(\vek w-\vek p_1,\vek p_1)}{D(\vek w)\xi(\vek w-\vek p_2,\vek p_2)}{}\overline{\hat\f(\vek w-\vek p_1,\vek p_1)}\hat\psi(\vek w-\vek p_2,\vek p_2)\,\times\nonumber\\
&&\times d^n\vek p_1\,d^n\vek p_2\,d^n\vek w\,d^n\vek x=0
\end{eqnarray}
for all $\f,\,\psi\in C_c(\nR^{2n})$.
\end{prop}

Choosing the constant kernel $\alpha\in\mf A(\nR^n)$, $\alpha(\vek w,\vek p_1,\vek p_2)=1$ for a.a.\ $\vek w,\,\vek p_1,\,\vek p_2\in\nR^n$, in \eqref{prae} one obtains an observable which we call as the {\it canonical position difference observable} $E_\ka$. A simple calculation shows that
\begin{equation}\label{ekan}
(E_\ka(B)\f)(\vek x,\vek y)=\chi_B(\vek y-\vek x)\f(\vek x,\vek y)
\end{equation}
for all $B\in\mc B(\nR^n)$, $\f\in L^2(\nR^{2n})$ and a.a.\ $\vek x,\,\vek y\in\nR^n$.

We may add another requirement for a position difference observable. Consider a representation $V_0$ of $\nR^n$ in $L^2(\nR^{2n})$ defined through
$$
\big(V_0(\vek p)\f\big)(\vek x,\vek y)=e^{i(\vek p|\vek x+\vek y)}\f(\vek x,\vek y)
$$
for all $\vek p,\,\vek x,\,\vek y\in\nR^n$ and $\f\in L^2(\nR^{2n})$. This representation corresponds to equal momentum boosts for both particles. We denote the set of those $M\in\mc O_U\big(\nR^n,\,L^2(\nR^{2n})\big)$ invariant with respect to $V_0$ by $\mf D(\nR^n)$, i.e.\ for all $M\in\mf D(\nR^n)$
\begin{equation}\label{mominv2}
V_0(\vek p)M(B)V_0(\vek p)^*=M(B),\qquad\vek p\in\nR^n,\quad B\in\mc B(\nR^n).
\end{equation}
Suppose that $M$ is as in \eqref{prae}. One can easily check that \eqref{mominv2} is equivalent with the condition
$$
\alpha(\vek w+2\vek p,\vek p_1+\vek p,\vek p_2+\vek p)=\alpha(\vek w,\vek p_1,\vek p_2)
$$
for a.a.\ $\vek w,\,\vek p_1,\,\vek p_2,\,\vek p\in\nR^n$. Let us denote the (essentially) constant value of the function $\nR^n\ni\vek p\mapsto\alpha(\vek w+2\vek p,\vek p_1+\vek p,\vek p_2+\vek p)$ by $\beta(\vek p_1-\vek w/2,\vek p_2-\vek w/2)$ for all $\vek w,\,\vek p_1,\,\vek p_2\in\nR^n$. Thus the function $\beta:\nR^{2n}\to\nC$ is the kernel of a positive form on $C_c(\nR^n)$ with diagonal values $\beta^{(0)}(\vek p)=1$ for a.a.\ $\vek p\in\nR^n$ and there is a weakly measurable unit-vector-valued function $\zeta:\nR^n\to L^2(\nR^n)$ such that
$$
\beta(\vek p_1,\vek p_2)=\sis{\zeta(\vek p_1)}{\zeta(\vek p_2)}{},\qquad\vek p_1,\,\vek p_2\in\nR^n.
$$
We conclude that for any $M\in\mf D(\nR^n)$ there is a weakly measurable unit vector valued map $\zeta:\nR^n\to L^2(\nR^n)$ such that
\begin{eqnarray}
\sis{\f}{M(B)\psi}{}&=&(2\pi)^{-n}\int_B\int_{\nR^n}\int_{\nR^n}\int_{\nR^n}e^{i(\vek x|\vek p_1-\vek p_2)}\sis{\zeta(\vek p_1-\vek w/2)}{\zeta(\vek p_2-\vek w/2)}{}\overline{\hat\f(\vek w-\vek p_1,\vek p_1)}\times\nonumber\\
&\times&\hat\psi(\vek w-\vek p_2,\vek p_2)\,d^n\vek p_1\,d^n\vek p_2\,d^n\vek w\,d^n\vek x
\end{eqnarray}
for all $\f,\,\psi\in C_c(\nR^{2n})$ and $B\in\mc B(\nR^n)$.

We may also demand invariance under non-equal momentum boosts for the particles, i.e.\ invariance with respect to the representation $V$ of $\nR^{2n}$ in $L^2(\nR^{2n})$ defined as the representation $U_{\nR^{2n}}$ of Section \ref{cpo}. It is an easy task to check that if we demand $V$-invariance of an observable $M\in\mc O_U\big(\nR^n,\,L^2(\nR^{2n})\big)$ defined by \eqref{prae}, we obtain the condition
$$
\alpha(\vek w+\vek p+\vek p',\vek p_1+\vek p',\vek p_2+\vek p')=\alpha(\vek w,\vek p_1,\vek p_2)
$$
for a.a.\ $\vek w,\,\vek p_1,\,\vek p_2,\,\vek p,\,\vek p'\in\nR^n$. We may thus replace the kernel $\alpha$ with a continuous function of positive type $\eta\in\mf E(\nR^n)$. This means that
\begin{eqnarray}
\sis{\f}{M(B)\psi}{}&=&(2\pi)^{-n}\int_B\int_{\nR^n}\int_{\nR^n}\int_{\nR^n}e^{i(\vek x|\vek p_1-\vek p_2)}\eta(\vek p_1-\vek p_2)\overline{\hat\f(\vek w-\vek p_1,\vek p_1)}\hat\psi(\vek w-\vek p_2,\vek p_2)\times\nonumber\\
&\times&d^n\vek p_1\,d^n\vek p_2\,d^n\vek w\,d^n\vek x
\end{eqnarray}
for all $\f,\,\psi\in C_c(\nR^{2n})$ and $B\in\mc B(\nR^n)$. Let $E_\ka$ be the canonical position difference observable of \eqref{ekan}. Denote the scalar measure $B\mapsto\sis{\f}{E_\ka(B)\psi}{}$ by $E_\ka^{\f,\psi}$. Using Bochner theorem one may easily check that $M=\rho*E_\ka$ with a probability measure $\rho:\mc B(\nR^n)\to[0,1]$, i.e.
$$
\sis{\f}{M(B)\psi}{}=(\rho*E_\ka^{\f,\psi})(B),\qquad\f,\,\psi\in L^2(\nR^{2n}).
$$
The converse also holds: $\rho*E_\ka$ is a $V$-invariant element in $\mc O_U\big(\nR^n,\,L^2(\nR^{2n})\big)$. Again, the extreme points in this smaller set of position difference observables are exactly the translated canonical position difference observables $E_{\vek u}$, $E_{\vek u}(B)=E_\ka(B-\vek u)$ for all $B\in\mc B(\nR^n)$. Next we consider the case of a compact cyclic position space.

Let a unitary representation $U$ of $\nT^{2n}$ in $L^2(\nT^{2n},\mu_{2n})$  be defined by
$$
\big(U(\vek u)\f\big)(\vek v)=\f(\vek v\overline{\vek u})
$$
for all $\vek u,\,\vek v\in\nT^{2n}$ and $\f\in L^2(\nT^{2n},\mu_{2n})$ where the measure $\mu_{2n}$ is defined as in the previous section. The physical system consists of two particles confined to the space $\nT^n$. The position difference of the particle 1 with coordinates $\vek w$ and the particle 2 with coordinates $\vek z$ is $\vek z\overline{\vek w}$. The Hilbert space describing the system is $L^2(\nT^{2n},\mu_{2n})$. Analogously to the above situation we define the {\it position difference observables} $M$ on $\nT^n$ through the covariance condition
$$
U(\vek w,\vek z)M(B)U(\vek w,\vek z)^*=M(\vek z\overline{\vek w}B)
$$
for all $B\in\mc B(\nT^n)$ and $\vek w,\,\vek z\in\nT^n$.

Define the orthonormal basis $\{e_{\vek m}\,|\,\vek m\in\nZ^{2n}\}$ where $e_{\vek m}(\vek u)=\overline{\la\vek u,\vek m\ra}$ for all $\vek m\in\nT^{2n}$ and $\vek u\in\nT^{2n}$. We may characterize the set of position difference observables $M\in\mc O_U\big(\nT^n,\,L^2(\nT^{2n})\big)$ and its extreme points in this situation in a similar fashion as above.
\begin{prop}
For any $M\in\mc O_U\big(\nT^n,\,L^2(\nT^{2n})\big)$ there is a family $\{\xi_{\vek k,\vek l}\,|\,\vek k,\,\vek l\in\nZ^n\}$ of unit vectors in an infinite dimensional Hilbert space $\mc M$ such that
\begin{equation}\label{torusero}
M(B)=\sum_{\vek j,\,\vek k,\,\vek l\in\nZ^n}\sis{\xi_{\vek j-\vek k,\vek k}}{\xi_{\vek j-\vek l,\vek l}}{}\int_B\la\vek z,\vek k-\vek l\ra\,d\mu_{n}(\vek z)|e_{\vek j-\vek k,\vek k}\ra\la e_{\vek j-\vek l,\vek l}|,\qquad B\in\mc B(\nT^n).
\end{equation}
On the other hand, given any family of unit vectors $\xi_{\vek k,\vek l}\in\mc M$, $\vek k,\,\vek l\in\nZ^n$, \eqref{torusero} defines a position difference observable $M\in\mc O_U\big(\nT^n,\,L^2(\nT^{2n})\big)$. Suppose that a position difference observable $M$ is as in \eqref{torusero}. Denote the Hilbert space generated by the vectors $\xi_{\vek j-\vek k,\vek k}$, $\vek k\in\nZ^n$, by $\vek H_{\vek j}$ for every $\vek j\in\nZ^n$. Define also $\vek H:=\bigoplus_{\vek j\in\nZ^n}\vek H_{\vek j}$. The observable $M$ is extreme in $\mc O_U\big(\nT^n,\,L^2(\nT^{2n})\big)$ if and only if there is no non-zero decomposable operator $A\in\mc L(\vek H)$ with components $A_{\vek j}\in\vek H_{\vek j}$, $\vek j\in\nZ^n$, such that
\begin{equation}
\sis{\xi_{\vek j-\vek k,\vek k}}{A_{\vek j}\xi_{\vek j-\vek k,\vek k}}{}=0
\end{equation}
for all $\vek j,\,\vek k\in\nZ^n$. The observable $M$ is extreme in $\mc O\big(\nT^n,\,L^2(\nT^{2n},\mu_{2n})\big)$ if and only if there is no non-zero decomposable operator $D\in\mc L\big(L^2(\nT^n,\mu_n;\vek H)\big)$ with components $D(\vek z)\in\mc L(\vek H)$, $\vek z\in\nT^n$, such that
\begin{equation}
\int_{\nT^n}\la\vek z,\vek k-\vek l\ra\sis{\xi_{\vek j-\vek k,\vek k}}{D(\vek z)\xi_{\vek j-\vek l,\vek l}}{}\,d\mu_n(\vek z)=0
\end{equation}
for all $\vek j,\,\vek k,\,\vek l\in\nZ^n$. The observable $M$ is a PVM if and only if the vectors $\nT^n\times\nZ^n\ni(\vek z,\vek j)\mapsto\la\vek z,\vek k\ra\xi_{\vek j-\vek k,\vek k}$, $\vek k\in\nZ^n$, generate the space $L^2(\nT^n,\mu_n;\vek H)$.
\end{prop}

Choosing in \eqref{torusero} $\xi_{\vek k,\,\vek l}=\xi\in\mc M$ for all $\vek k,\,\vek l\in\nZ^n$, one obtains the {\it canonical position difference observable} $F_\ka$, $\big(F_\ka(B)\f\big)(\vek w,\vek z)=\chi_B(\vek z\overline{\vek w})\f(\vek w,\vek z)$ for all $B\in\mc B(\nZ^n)$, $\f\in L^2(\nT^{2n},\mu_{2n})$ and $\vek w,\,\vek z\in\nT^n$. This is, of course a PVM. We may again require additional invariance properties of position difference observables on $\nT^n$. One can proceed as earlier in this section and demand different momentum boost invariance conditions; note that in space $\nT^n$ the momentum is quantized. We obtain results analogous to those above and we will not go into them in detail. Again, demanding invariance under independent momentum boosts for the particles, one ends up with a convolution structure: any such position difference observable is of the form $\rho*F_\ka$ with some probability measure $\rho:\mc B(\nT^n)\to[0,1]$. Note that restricting the momentum space, one can define {\it covariant phase difference observables} \cite{hei} and get similar results.

\section*{Conclusions}

We have characterized the quantum observables, POVMs, which are extreme in the set of observables covariant with respect to a unitary representation of a locally compact Abelian group which is Hausdorff in the case of an arbitrary transitive value space. The method can be applied to several physical situations of which time, position and position difference observables are treated. We have also characterized the covariant observables which are extreme in the set of all quantum observables.

The corresponding results concerning a very general type I symmetry group $G$ with the trivial value space $(G,\,\mc B(G))$ have been obtained in \cite{ho3,ho1} and in the case of a compact symmetry group and an arbitrary transitive value space in \cite{car}. The next task is to generalize the results of this paper. An aim is to characterize the covariance structures involving a more general type I group and an arbitrary transitive value space and study covariant extreme observables using the methods presented especially in \cite{cas} and \cite{ho1}. This would allow us to study a wider range of covariant observables and give their extremality conditions. Especially, interesting cases would be extreme covariant observables with Euclidean, Poincar\'e or Heisenberg symmetry groups.

\section*{Acknowledgements}

The authors would like to thank Pekka Lahti for previewing this paper. The work is supported by The Academy of Finland grant no.\ 138135.

\appendix

\section{Appendix: Proofs of Theorems \ref{theor1}, \ref{theor2}, and \ref{theor4}}\label{aa}

Before we prove Theorem \ref{theor1}, we give the exhaustive characterization of the covariance structure $\mc O_U(\Om,\,\hil)$ found in \cite{cas}. Suppose that $M\in\mc O_U(\Om,\,\hil)$ and that $f:\Om\to\nC$ is continuous and compactly supported. We may now define a bounded operator $M(f)\in\mc L(\hil)$ by
$$
M(f)=\int_\Om f(\om)\,dM(\om).
$$
Let us state the main result of \cite{cas}.

\begin{theor}\label{castheor}
Suppose that $\mc O_U(\Om,\,\hil)\neq\emptyset$. Let the measure $\tilde\nu$ and the spaces $\hil_\gamma$, $\gamma\in\hat G$, be as in Section \ref{cdt}. Suppose that $M\in\mc O_U(\Om,\,\hil)$ and that $\mc M$ is a fixed infinite-dimensional Hilbert space. There is a decomposable isometry $W:\hil\to L^2(\hat G,\tilde\nu;\mc M)$ such that $(W\f)(\gamma)=W(\gamma)\f(\gamma)$ for all $\f\in\hil$, where $W(\gamma):\hil_\gamma\to\mc M$ is an isometry for $\tilde\nu$-a.a.\ $\gamma\in\hat G$. Furthermore for all $\f,\,\psi\in\hil$ and $f\in C_c(\Om)$ one has
\begin{equation}\label{caskaa}
\sis{\f}{M(f)\psi}{}=\int_{\hat G}\int_{H^\perp}(\mc F f)(\eta)\sis{\f(\gamma)}{W(\gamma)^*W(\gamma-\eta)\psi(\gamma-\eta)}{}\,d\eta\,d\tilde\nu(\gamma).
\end{equation}
Conversely, if $\mc M$ is an infinite dimensional Hilbert space and $W$ a decomposable isometry, \eqref{caskaa} defines an observable $M\in\mc O_U(\Om,\,\hil)$.
\end{theor}

\begin{rem}\label{tih}\rm
If we do not demand $\nu_U=\tilde\nu$ and only assume that $d\nu_U(\gamma)=\rho(\gamma)\,d\tilde\nu(\gamma)$ with a real non-negative Borel-function $\rho$ on $\hat G$ we must replace the isometric components $W(\gamma)$ of $W$ by $\sqrt{\rho(\gamma)}W(\gamma)$ in the above theorem as well as in Theorems \ref{theor1}, \ref{theor2} and \ref{theor4}. This is unnecessary since the density function $\rho$ can be embedded in the component spaces $\hil_\gamma$ of the decomposition \eqref{suora} for $\hil$ and thus we may assume with no loss of generality that $\rho=1$.
\end{rem}

Next we reformulate the above theorem to obtain Theorem \ref{theor1}. Fix an infinite-dimensional Hilbert space $\mc M$ and let $W:\hil\to L^2(\hat G,\,\tilde\nu;\,\mc M)$ be a decomposable isometry so that $(W\f)(\gamma)=W(\gamma)\f(\gamma)$ for all $\f\in\hil$ and $\gamma\in\hat G$ where $W(\gamma):\hil_\gamma\to\mc M$ is an isometry and let the operator $\mf W$ and spaces $\mc D\subset\hil$ and $\vek H$ be as in Section \ref{cdt}. First we show that $\mc D$ is dense in $\hil$. Let us denote the measurable function $\gamma\mapsto\dim{(\hil_\gamma)}$ by $n$. Suppose that $\hat G\ni\gamma\mapsto\{e_k(\gamma)\in\hil_\gamma\,|\,1\leq k<n(\gamma)+1\}$ is a measurable field of orthonormal bases in the component spaces $\hil_\gamma$, $\gamma\in\hat G$. This means that every $\f\in\hil$ can be written in the form
$$
\f(\gamma)=\sum_{k=1}^{n(\gamma)}\sis{e_k(\gamma)}{\f(\gamma)}{}e_k(\gamma),\qquad\gamma\in\hat G.
$$
We associate with every $\f\in\hil$ the functions $\f_k:\hat G\to\nC$, $k\in\nN\cup\{\infty\}$, defining for all $\gamma\in\hat G$ $\f_k(\gamma)=\sis{e_k(\gamma)}{\f(\gamma)}{}$ for all $k<n(\gamma)+1$ and otherwise $\f_k(\gamma)=0$. We denote by ${\mc D}_0 $ the dense subspace of vectors $\f\in\hil$ such that $\f_k\in C_c(\hat G)$ for all $k\in\nN\cup\{\infty\}$ and only a finite number of the functions $\f_k$ differ from zero. One immediately sees that ${\mc D}_0 \subset\mc D$ and thus $\mc D$ is also dense. Note that $\hil$ can be embedded in $L^2(\hat G,\tilde\nu;\mathcal M)$ via $\sum_{k=1}^{n(\gamma)}\f_k(\gamma)e_k(\gamma)\mapsto\sum_{k=1}^{\infty}\f_k(\gamma)f_k$ where $\{f_k\}_{k=1}^\infty$ is an orthonormal basis of $\mathcal M$.

\emph{Proof of Theorem \ref{theor1}}: Suppose that $M\in\mc O_U(\Om,\,\hil)$ and that the decomposable isometry $W$ is as in Theorem \ref{castheor}. Let the operator $\mf W$ with domain $\mc D$ and image space closure $\vek H$ be as earlier. When $\f,\,\psi\in\hil$ and $f\in C_c(\Om)$ we may rewrite \eqref{caskaa} in the following way using the properties of $\tilde\nu$:
\begin{eqnarray}
\sis{\f}{M(f)\psi}{}&=&\int_{\hat G/H^\perp}\int_{H^\perp}\int_{H^\perp}(\mc Ff)(\eta)\sis{W(\gamma+\zeta)\f(\gamma+\zeta)}{W(\gamma+\zeta-\eta)\psi(\gamma+\zeta-\eta)}{}\,d\eta\,d\zeta\,d\nu([\gamma])\nonumber\\
&=&\int_{\hat G/H^\perp}\int_{H^\perp}\int_{H^\perp}(\mc Ff)(\zeta-\xi)\sis{W(\gamma+\zeta)\f(\gamma+\zeta)}{W(\gamma+\xi)\psi(\gamma+\xi)}{}\,d\xi\,d\zeta\,d\nu([\gamma])\nonumber\\
&=&\int_{\hat G/H^\perp}\int_{H^\perp}\int_{H^\perp}\int_\Om\la g,\zeta-\xi\ra f([g])\sis{W(\gamma+\zeta)\f(\gamma+\zeta)}{W(\gamma+\xi)\psi(\gamma+\xi)}{}\times\nonumber\\
&\times&d\mu([g])\,d\xi\,d\zeta\,d\nu([\gamma])\nonumber\\
&=&\int_{\hat G/H^\perp}\int_{H^\perp}\int_{H^\perp}\int_\Om f([g])\sis{W(\gamma+\zeta)\big(U(g)^*\f\big)(\gamma+\zeta)}{W(\gamma+\xi)\big(U(g)^*\psi\big)(\gamma+\xi)}{}\times\nonumber\\
&\times&d\mu([g])\,d\xi\,d\zeta\,d\nu([\gamma])\nonumber
\end{eqnarray}
giving the first part of Theorem \ref{theor1}

Suppose now that there is a projection valued measure $P\in\mc O_U(\Om,\,\hil)$. We may proceed as in \cite{hoa} and define operators $V(\eta)\in\mc L(\hil)$, $\eta\in H^\perp$, through
$$
V(\eta)=\int_\Om\la g,\eta\ra\,dP([g]).
$$
Since $P$ is projection valued it follows that $\eta\mapsto V(\eta)$ is a unitary representation of $H^\perp$ \cite{hoa}. As in \cite[Section 5]{hoa} one can calculate using Fourier-Plancherel theory that
$$
\big(V(\eta)\f\big)(\gamma)=W(\gamma)^*W(\gamma-\eta)\f(\gamma-\eta)
$$
for a.a. $\gamma\in\hat G$, $\eta\in H^\perp$ and all $\f\in\hil$ where $W$ is the decomposable isometry defining $P$. Using the above equation and $V(\zeta)V(\xi)=V(\zeta+\xi)$ one finds that
$$
W(\gamma)^*W(\gamma-\zeta-\xi)=W(\gamma)^*W(\gamma-\zeta)W(\gamma-\zeta)^*W(\gamma-\zeta-\xi)
$$
for a.a. $\gamma\in\hat G$ and $\zeta,\,\xi\in H^\perp$. Substituting $\xi=-\zeta$ in the above formula we find
$$
I_{\hil_\gamma}=W(\gamma)^*W(\gamma-\zeta)W(\gamma-\zeta)^*W(\gamma)=W(\gamma)^*W(\gamma-\zeta)\big(W(\gamma)^*W(\gamma-\zeta)\big)^*.
$$
Replacing $\gamma$ with $\gamma+\zeta$ produces
$$
I_{\hil_{\gamma+\zeta}}=W(\gamma+\zeta)^*W(\gamma)W(\gamma)^*W(\gamma+\zeta)=\big(W(\gamma)^*W(\gamma+\zeta)\big)^*W(\gamma)^*W(\gamma+\zeta).
$$
Thus $W(\gamma_2)^*W(\gamma_1):\hil_{\gamma_1}\to\hil_{\gamma_2}$ is unitary for a.a.\ $\gamma_1,\,\gamma_2$ within each coset $[\gamma]\in\hat G/H^\perp$. It follows that (almost) all $\hil_\gamma$ within the same coset have the same dimension. We also see immediately that the image spaces $W(\gamma)(\hil_\gamma)$ are equal within each coset of $\hat G/H^\perp$. On the other hand if the mapping $\gamma\mapsto\dim{(\hil_\gamma)}$ is essentially constant on a.a.\ $[\gamma]\in\hat G/H^\perp$ one may easily construct a decomposable isometry $W$ with the above property and thus a PVM. $\Box$

Before proving Theorem \ref{theor2} we show that the space $\vek H=\overline{\mf W(\mc D)}$ (where the operator $\mf W:\mc D\to L^2(\hat G/H^\perp,\nu;\,\mc M)$ is constructed from a decomposable isometry as in Section \ref{cdt}) has the direct integral decomposition \eqref{suorah}. Suppose that $P_{\vek H}$ is the projection of $L^2(\hat G/H^\perp,\,\nu;\,\mc M)$ onto $\vek H$. Define also two PVMs $P\in\Sigma\big(\hat G/H^\perp,\,L^2(\hat G/H^\perp,\nu;\mc M)\big)$ and $E\in\Sigma(\hat G/H^\perp,\,\hil)$ by setting
$$
P(B_1)\F=\chi_{B_1}\F,\qquad E(B_2)\f=(\chi_{B_2}\circ p)\f
$$
for all $B_1,\,B_2\in\mc B(\hat G/H^\perp)$, $\F\in L^2(\hat G/H^\perp,\,\nu;\,\mc M)$ and $\f\in\hil$. Here $p:\hat G\to\hat G/H^\perp$ is the canonical quotient mapping. We immediately note that
$$
P(B)\mf W\f=\mf WE(B)\f
$$
for all $\f\in\mc D$. Since $E(B)\f\in\mc D$ whenever $B\in\mc B(\hat G/H^\perp)$ and $\f\in\mc D$ it follows that $P(B)(\vek H)\subset\vek H$ for all $B\in\mc B(\hat G/H^\perp)$. This implies that also $P(B)(\vek H^\perp)\subset\vek H^\perp$ for all $B\in\mc B(\hat G/H^\perp)$. One easily finds that
$$
[P(B),P_{\vek H}]=0
$$
for all $B\in\mc B(\hat G/H^\perp)$. This means that the projection $P_{\vek H}$ is decomposable, i.e.\ there is a weakly measurable field $\hat G/H^\perp\ni[\gamma]\mapsto P_{[\gamma]}\in\mc L(\mc M)$ of projections such that $(P_{\vek H}\F)([\gamma])=P_{[\gamma]}\F([\gamma])$ for all $\F\in L^2(\hat G/H^\perp,\,\nu;\,\mc M)$ and $[\gamma]\in\hat G/H^\perp$. Denote the image spaces $P_{[\gamma]}(\mc M)$ by $\vek H_{[\gamma]}$ for all $[\gamma]\in\hat G/H^\perp$. This verifies the decomposition \eqref{suorah}.

\emph{Proof of Theorem \ref{theor2}:} Suppose that $A\in\mc L(\vek H)$ is a non-zero decomposable operator such that \eqref{extehto} holds. We may assume that $\|A\|\leq1$; otherwise one can redefine $A'=\|A\|^{-1}A$. With no loss of generality we may also assume that $A$ is selfadjoint; otherwise redefine $A''=i(A-A^*)$. Let us define positive (decomposable) operators $A^\pm=I\pm A$ and operator-valued measures $M^\pm$ by
$$
\sis{\f}{M^\pm(B)\psi}{}=\int_B\int_{\hat G/H^\perp}\int_{H^\perp}\int_{H^\perp}\la g,\zeta-\xi\ra\sis{W(\gamma+\zeta)\f(\gamma+\zeta)}{A^\pm_{[\gamma]}W(\gamma+\xi)\psi(\gamma+\xi)}{}\,d\zeta\,d\xi\,d\nu([\gamma])\,d\mu([g])
$$
for all $\f,\,\psi\in\mc D$ and $B\in\mc B(\Om)$. Using Fourier-Plancherel theory, Equation \eqref{extehto} and properties of $\tilde\nu$, we have
\begin{eqnarray}
\sis{\f}{M^\pm(\Om)\psi}{}&=&\int_\Om\int_{\hat G/H^\perp}\int_{H^\perp}\int_{H^\perp}\la\om,\zeta-\xi\ra\sis{W(\gamma+\zeta)\f(\gamma+\zeta)}{A^\pm_{[\gamma]}W(\gamma+\xi)\psi(\gamma+\xi)}{}\times\nonumber\\
&\times&d\zeta\,d\xi\,d\nu([\gamma])\,d\mu(\om)\nonumber\\
&=&\int_{\hat G/H^\perp}\int_{H^\perp}\sis{W(\gamma+\eta)\f(\gamma+\eta)}{A^\pm_{[\gamma]}W(\gamma+\eta)\psi(\gamma+\eta)}{}\,d\eta\,d\nu([\gamma])\nonumber\\
&=&\int_{\hat G/H^\perp}\int_{H^\perp}\sis{W(\gamma+\eta)\f(\gamma+\eta)}{W(\gamma+\eta)\psi(\gamma+\eta)}{}\,d\eta\,d\nu([\gamma])\nonumber\\
&=&\int_{\hat G}\sis{W(\gamma)\f(\gamma)}{W(\gamma)\psi(\gamma)}{}\,d\tilde\nu(\gamma)=\int_{\hat G}\sis{\f(\gamma)}{\psi(\gamma)}{}\,d\tilde\nu(\gamma)=\sis{\f}{\psi}{}.\nonumber
\end{eqnarray}
Thus $M^\pm$ are POVMs. In the above calculation we have identified $\hat\Om$ with $H^\perp$. It follows from the decomposability of $A$ that $M^\pm\in\mc O_U(\Om,\,\hil)$. Since $A\neq 0$ we have $M^+\neq M^-$ and $M=\frac12(M^++M^-)$ and hence $M\notin\mr{Ext}(\mc O_U(\Om,\,\hil))$.

Let us now assume that $M\notin\mr{Ext}(\mc O_U(\Om,\,\hil))$ so that there are distinct covariant observables $M^\pm\in\mc O_U(\Om,\,\hil)$ such that $M=\frac12(M^++M^-)$. The isometries of Theorem \ref{castheor} related to $M^\pm$ are denoted by $W^\pm$. Let the corresponding operators giving the Kolmogorov decompositions \eqref{kolm} be $\mf W^\pm$ and let the Hilbert space completions of the images $\mf W^\pm(\mc D)$ be $\vek H^\pm$ with components $\vek H_{[\gamma]}^\pm$, $[\gamma]\in\hat G/H^\perp$, respectively. Suppose that $\f,\,\psi\in\mc D$ are such that $\mf W\f=\mf W\psi$. Since $\|\mf W^\pm\chi\|\leq\sqrt 2\|\mf W\chi\|$ for all $\chi\in\mc D$, we have $\mf W^\pm(\f-\psi)=0$. Thus one may define sesquilinear forms $F^\pm:\vek H\times\vek H\to\nC$ by extending 
$$
F^\pm(\mf W\f,\mf W\psi)=\sis{\mf W^\pm\f}{\mf W^\pm\psi}{},\;\;\;\; \f,\,\psi\in\mc D,
$$
to $\vek H\times\vek H$. We now have $F^\pm(\F,\F)\leq2\|\F\|^2$ for all $\F\in\vek H$. Hence the forms $F^\pm$ are well defined and bounded and there are operators $A^\pm\in\mc L(\vek H)$ such that $F^\pm(\F,\Psi)=\sis{\F}{A^\pm\Psi}{\vek H}$. It follows that
\begin{equation}\label{ops}
\sis{\f}{M^\pm(B)\psi}{}=\int_B\sis{\mf W(U\circ s)(\om)^*\f}{A^\pm\mf W(U\circ s)(\om)^*\psi}{}\,d\mu(\om)
\end{equation}
for all $\f,\,\psi\in\mc D$, $B\in\mc B(\Om)$ and any Borel measurable section $s:\Om\to G$; such sections exist \cite[lemma 3]{bag}.

Define a unitary representation $\Pi$ of $H$ in the Hilbert space $\vek H$ by
\begin{equation}\label{lambda}
\big(\Pi(h)\F\big)([\gamma])=\la h,\gamma\ra\F([\gamma])
\end{equation}
for all $\F\in\vek H$, $h\in H$ and $[\gamma]\in\hat G/H^\perp$. One has
\begin{equation}\label{lambdakom}
\mf WU(h)\f=\Pi(h)\mf W\f
\end{equation}
for all $h\in H$ and $\f\in\mc D$.

Using \eqref{lambdakom} the arbitrariness of the section $s$ means that $A^\pm$ must commute with the representation $\Pi$. This means that $A^\pm$ are decomposable as in \eqref{suoraA} \cite{dix}. Since $M^+\neq M^-$ one must have $A^+\neq A^-$. We may now define a nonzero decomposable operator $A=A^+-A^-\in\mc L(\vek H)$ with the property
$$
\int_\Om\sis{U(g)^*\f}{\mf W^*A\mf WU(g)^*\psi}{}\,d\mu([g])=\sis{\f}{\psi}{}-\sis{\f}{\psi}{}=0
$$
for all $\f,\,\psi\in\mc D$. Applying Fourier-Plancherel theory to the above equation one finds that
\begin{eqnarray}
0&=&\int_\Om\int_{\hat G/H^\perp}\int_{H^\perp}\int_{H^\perp}\la g,\zeta-\xi\ra\sis{W(\gamma+\zeta)\f(\gamma+\zeta)}{A_{[\gamma]}W(\gamma+\xi)\psi(\gamma+\xi)}{}\,d\zeta\,d\xi\,d\nu([\gamma])\,d\mu([g])\nonumber\\
&=&\int_{\hat G/H^\perp}\int_{H^\perp}\sis{W(\gamma+\zeta)(\f(\gamma+\zeta)}{A_{[\gamma]}W(\gamma+\zeta)\psi(\gamma+\zeta)}{}\,d\zeta\,d\nu([\gamma])\nonumber
\end{eqnarray}
for all $\f,\,\psi\in\mc D$. This means that $A$ satisfies the condition \eqref{extehto}.
$\Box$

\emph{Proof of Theorem \ref{theor4}:} Suppose that $M\in\mc O_U(\Om,\,\hil)$. Let us pick a measurable section $s:\Om\to G$. Using the notations of Section \ref{cdt} we may define an isometry $\mc J_s:\hil\to L^2(\Om,\mu;\vek H)$ by
\begin{equation}\label{jisom}
(\mc J_s\f)(\om)=\mf W(U\circ s)(\om)^*\f
\end{equation}
for all $\f\in\mc D$ and $\om\in\Om$. Let us also define a PVM $P\in\Sigma(\mc B(\Om),\,L^2(\Om,\,\mu;\,\vek H))$ by $P(B)\psi=\chi_B\psi$ for all $B\in\mc B(\Om)$ and $\psi\in L^2(\Om,\mu;\vek H)$. It follows from the definition of the space $\vek H$ that vectors $\mf W\f$, $\f\in\mc D$, form a dense subspace of $\vek H$. It now immediately follows that vectors of the form $\chi_B(\cdot)\mf W(U\circ s)(\cdot)^*\f$ where $B\in\mc B(\Om)$ and $\f\in\mc D$, form a dense subset of $L^2(\Om,\,\mu;\,\vek H)$. Hence the triple $\big(L^2(\Om,\mu;\vek H),\,P,\,\mc J_s\big)$ is the minimal Naimark dilation of $M$, i.e.\ $M(B)=\mc J_s^*P(B)\mc J_s$ or
$$
\sis{\f}{M(B)\psi}{}=\int_B\sis{(\mc J_s\f)(\om)}{(\mc J_s\psi)(\om)}{}\,d\mu(\om)
$$
for all $\f,\,\psi\in\mc D$ and $B\in\mc B(\Om)$.

The Naimark dilation of the above form allows us to investigate whether the covariant observable $M$ is extreme in $\mc O(\Om,\,\hil)$. In \cite{hyt} Hyt\"onen {\it et al} found a `diagonal' minimal Naimark dilation $(\mc K,\,P,\,\mc J)$ for any $M\in\mc O(\mc A,\,\hil)$ with an arbitrary measurable space $(\Om,\,\mc A)$ where $\mc K$ is of the direct integral form $\suor{\Om}\mc K_\om\,dm(\om)$ and the spectral measure $P\in\Sigma(\mc A,\,\mc K)$ is defined through $P(B)\F=\chi_B\F$ for all $\F\in\mc K$ and $B\in\mc A$. The ($\sigma$-)finite measure $m$ here is such that the measures $B\mapsto\sis{\f}{M(B)\psi}{}$ are absolutely continuous with respect to $m$. (Such measures always exist as shown in \cite{hyt}). The extreme points of $\mc O(\mc A,\,\hil)$ were characterized in \cite{pel} using this special form of the minimal Naimark dilation. Since $G$ is locally compact and second countable it is also $\sigma$-compact and thus $\mu$ is $\sigma$-finite. The measures $B\mapsto\sis{\f}{M(B)\psi}{}$, $\f,\,\psi\in\hil$, are absolutely continuous with respect to $\mu$ as we have seen in Theorem \ref{theor1}. Thus the dilation obtained above for a covariant observable $M$ is of this special diagonal form and we may readily use the results of \cite{pel}.

The requirement for a covariant observable $M\in\mc O_U(\Om,\,\hil)$ with $\mc J_s$ to be extreme in $\mc O(\Om,\,\hil)$ is the following \cite{pel}: if $D\in\mc L\big(L^2(\Om,\,\mu;\,\vek H)\big)$ is decomposable, the condition $\mc J_s^*D\mc J_s=0$ implies $D=0$. Suppose that $s':\Om\to G$ is another measurable section and that $\mc J_{s'}$ is the corresponding isometry giving the unitarily equivalent Naimark dilation of $M$. Suppose also that $D\in\mc L\big(L^2(\Om,\,\mu;\,\vek H)\big)$ is decomposable with components $D(\om)\in\mc L(\vek H)$ for all $\om\in\Om$. One finds that
$$
\mc J_{s'}^*D\mc J_{s'}=\mc J_s^*\tilde D\mc J_s,
$$
where $\tilde D\in\mc L\big(L^2(\Om,\,\mu;\,\vek H)\big)$ is a decomposable operator with components $\tilde D(\om)=\Pi\big(h(\om)\big)D(\om)\Pi\big(h(\om)\big)^*$ for all $\om\in\Om$. Here $h:\Om\to H$, $h(\om)=s'(\om)s(\om)^{-1}$, $\om\in\Om$, is a measurable map. As a result we find that if $\mc J_s^*D\mc J_s=0$ for all decomposable $D\in\mc L\big(L^2(\Om,\,\mu;\,\vek H)\big)$ with some section $s$ the same true with any section. This concludes the proof of Theorem \ref{theor4}. $\Box$

\begin{rem}\rm
An observable $P\in\mc O_U(\Om,\,\hil)$ is a PVM if and only if the associated isometry $\mc J_s$ with any measurable section $s$ is unitary, i.e. $\mc J_s\mc J_s^*=I_{L^2(\Om,\,\mu;\,\vek H)}$. When we represent vectors $\F\in L^2(\Om,\,\mu;\,\vek H)$ in the form $\Om\times\hat G/H^\perp\ni(\om,[\gamma])\mapsto\F(\om,[\gamma])\in\vek H_{[\gamma]}$, a simple calculation shows that
\begin{equation}\label{viim}
(\mc J_s\mc J_s^*\F)(\om,[\gamma])=\int_{H^\perp}\int_\Om\la s(\om)^{-1}s(\om'),\gamma+\eta\ra E(\gamma+\eta)\F(\om',[\gamma])\,d\mu(\om')\,d\eta,
\end{equation}
where $E(\gamma):=W(\gamma)W(\gamma)^*\in\mc L(\vek H_{[\gamma]})$, $\gamma\in\hat G$, is the projection onto the image space $\overline{W(\gamma)(\hil_\gamma)}$, when $W$ is a decomposable isometry associated with $P$. As was shown in the proof of Theorem \ref{theor1}, these image spaces coincide within each coset of $\hat G/H^\perp$ and thus the projections $E(\gamma)$ coincide in each coset. We denote the relevant projections by $E_{[\gamma]}$, i.e. $E(\gamma')=E_{[\gamma]}$ whenever $\gamma'\in[\gamma]$. Applying this result to Equation \eqref{viim} gives
$$
(\mc J_s\mc J_s^*\F)(\om,[\gamma])=E_{[\gamma]}\F(\om,[\gamma])
$$
for a.a.\ $\om\in\Om$ and $[\gamma]\in\hat G/H^\perp$. If $P$ is a PVM it follows that $E_{[\gamma]}=I_{\vek H_{[\gamma]}}$ for a.a.\ $[\gamma]\in\hat G/H^\perp$. This means that $W(\gamma):\hil_\gamma\to\vek H_{[\gamma]}$ is unitary.
\end{rem}

\end{document}